\documentclass[twocolumn,american]{revtex4-1}
\usepackage[T1]{fontenc}
\usepackage[latin9]{inputenc}
\usepackage{babel}
\usepackage{amsmath}
\usepackage{amssymb}
\usepackage{graphicx}
\usepackage{esint}
\usepackage[unicode=true,
 bookmarks=true,bookmarksnumbered=false,bookmarksopen=false,
 breaklinks=true,pdfborder={0 0 0},backref=false,colorlinks=false]
 {hyperref}


\begin{document}

\title{Statistical Physics of Neural Systems\\
with Non-additive Dendritic Coupling}

\author{David Breuer$^{1,2,*}$, Marc Timme$^{1,3,4}$, and Raoul-Martin
Memmesheimer$^{5}$}

\affiliation{%
\mbox{%
$^{1}$Network Dynamics, Max Planck Institute for Dynamics and Self-Organization
(MPIDS), 37077 Göttingen, Germany%
}}

\affiliation{%
\mbox{%
$^{2}$Mathematical Modeling, Max Planck Institute for Molecular Plant
Physiology (MPIMP), 14476 Potsdam-Golm, Germany%
}}

\affiliation{%
\mbox{%
$^{3}$Bernstein Center for Computational Neuroscience (BCCN), 37077
Göttingen, Germany%
}}

\affiliation{%
\mbox{%
$^{4}$Institute for Nonlinear Dynamics, Faculty of Physics, Georg
August University, 37077 Göttingen, Germany%
}}

\affiliation{%
\mbox{%
$^{5}$Donders Institute, Department of Neuroinformatics, Radboud
University, 6525 AJ Nijmegen, Netherlands%
}}

\affiliation{$^{*}$breuer@mpimp-golm.mpg.de}
\begin{abstract}
How neurons process their inputs crucially determines the dynamics
of biological and artificial neural networks. In such neural and neural-like
systems, synaptic input is typically considered to be merely transmitted
linearly or sublinearly by the dendritic compartments. Yet, single-neuron
experiments report pronounced supralinear dendritic summation of sufficiently
synchronous and spatially close-by inputs. Here, we provide a statistical
physics approach to study the impact of such non-additive dendritic
processing on single neuron responses and the performance of associative
memory tasks in artificial neural networks. First, we compute the
effect of random input to a neuron incorporating nonlinear dendrites.
This approach is independent of the details of the neuronal dynamics.
Second, we use those results to study the impact of dendritic nonlinearities
on the network dynamics in a paradigmatic model for associative memory,
both numerically and analytically. We find that dendritic nonlinearities
maintain network convergence and increase the robustness of memory
performance against noise. Interestingly, an intermediate number of
dendritic branches is optimal for memory functionality.

~

Keywords: statistical physics, neural networks, associative memory,
nonlinear dendrites

~

PACS: 05.20.-y, 87.19.L-, 84.35.+i
\end{abstract}
\maketitle

\section{Introduction: Non-additive dendritic input processing in neural networks}

Information processing in artificial and biological neural networks
crucially depends on the processing of inputs in single neurons (e.g.~\cite{Koch2000}).
The dendrites, branched protrusions of a biological nerve cell or
the input preprocessing of formal neurons constitute the main input
sites. Traditionally, dendrites are modeled as passive, cable-like
conductors which integrate incoming presynaptic signals linearly or
sublinearly and propagate the change in voltage to the cell body or
soma where it is subject to nonlinear transformations \cite{stuart2007dendrites}.
Accordingly, the input preprocessing in formal neurons is usually
assumed to be a linear or sublinear summation.

Single-neuron experiments, however, demonstrate the occurrence of
strongly supralinear dendritic amplification. Biophysically, this
is caused by action potentials generated in the dendrite of the neuron.
Such dendritic spikes are mediated by voltage-dependent ion channels
such as sodium, calcium, and NMDA channels \cite{Ariav2003,Gasparini2004,Polsky2004,NLPS07,Larkum2008}.
In particular, dendritic spikes may emerge if sufficiently synchronous
inputs are received by the same branch of a dendrite. The many inputs
to the dendrites can thus be processed non-additively, depending on
their spatial and temporal distribution \cite{Poirazi2003,Polsky2004}.
This implies crucial deviations from the classical assumptions on
linear dendritic input processing as modeled, e.g., by cable equations.
It has been recently shown that dendritic spikes are present and prominent
all over the brain (e.g.~\cite{London2005}).

A number of theoretical studies highlighted the importance of nonlinear,
spiking dendrites already for the input processing in single neurons:
Simulations of neuron models with detailed channel density and morphology
showed dendritic spike generation in agreement with neurobiological
experiments \cite{Ariav2003,Poirazi2003a,Poirazi2003,Gasparini2004,NLPS07}.
Further, firing rate models have been developed \cite{Mel92a} which
reproduce the response properties of detailed models to diverse stimuli
and behave like multi-layered feed-forward networks of simple rate
neurons \cite{Poirazi2003a,Poirazi2003,Polsky2004}. Two and multi-layer
feed-forward networks of binary, deterministic neurons have been studied
using statistical physics methods \cite{Barkai1990,hertz1991introduction,Biehl2000}.
In particular, the so-called committee machine may be seen as a neuron
model incorporating a layer of dendrites with step-like activation
functions, i.e.~without analogous signal transmission \cite{Engel1992,Urbanczik1997}.
Neurons in biological networks receive time dependent, noisy input
at high rates which often makes a statistical description of the response
properties of single neurons necessary. In Ref.~\cite{ujfalussy2009parallel},
the authors derived such a description for linear and quadratic dendritic
summation together with some numerical results for a biologically
plausible, sigmoidal dendritic nonlinearity. The propagation of dendritic
spikes in branched dendrites with step-like activation functions has
been studied in Ref.~\cite{gollo2011statistical}, providing the
somatic input as a numerical solution to a high-dimensional system
of nonlinear equations. To date there is no efficient statistical
description for neurons with biologically plausible, sigmoidal dendritic
nonlinearities. In biological systems, neurons form complex, recurrent
networks. Thus, a description which allows to analytically study networks
of neurons with multiple nonlinear dendrites is especially desirable.

Recent single neuron experiments investigated the role of active dendrites
in detecting specific spatio-temporal input patterns \cite{Schiller2001,Gasparini2004,Losonczy2008}.
Theoretical studies showed that nonlinear dendrites improve the ability
of single neurons and ensembles of single neurons to discriminate
and learn different input patterns \cite{mel1992nmda,Poirazi2001,Rhodes2008,schiess2012gradient}.
Besides trivially multiplying the single neuron abilities to detect
input patterns, a\emph{ network} of neurons can store, retrieve and
complete spatio-temporal patterns aided by its recurrent dynamics:
It can function as an associative memory device \cite{Hopfield1982,Amit1985a,hertz1991introduction}.
Yet, the impact of nonlinear dendrites on associative memory networks
is unknown.

So far, only few studies considered the impact of non-additive dendrites
on network dynamics. Selectivity and invariance of network responses
to external stimuli and their intensity were analyzed in a firing
rate model \cite{Morita2007,Morita2008}. Refs.~\cite{Lisman1998,Wang2001,Morita2008}
proposed that NMDA-receptor dependent dendritic nonlinearities play
a crucial role in working memory, i.e., in the formation of persistent
activity in unstructured networks. Nonlinear, multiplicative dendritic
processing arising from spatial summation of input across the dendritic
arbor was similarly shown to enable spontaneous and persistent network
activity \cite{Zhang2013}. Dendritic spikes were suggested to work
as coincidence detectors and provide a neuronal basis for temporal
and spatial context in biological networks \cite{Katz2007,Stuart2001}.
Refs.~\cite{LJF10,TW82} studied networks of bursting neurons, where
the bursts facilitate the emergence of patterns of coordinated neuronal
activity and can be explained by dendritic spikes. Further, it was
shown that nonlinear dendrites can enable robust propagation of synchronous
spiking in random networks with biologically plausible substructures
\cite{jahnke2012guiding} and in purely random networks \cite{Memmesheimer2012}.
Finally, dendritic spikes were related to so-called sharp-wave ripples
in the hippocampus which are important for long-term memory consolidation
\cite{Memmesheimer2010}.

Networks of binary neurons with linear input summation have been intensively
investigated in statistical physics (``Hopfield networks'', \cite{Hopfield1982,Amit1985a,hertz1991introduction})
and extensions to different nonlinear and non-monotonic transfer functions
exist (cf.~e.g.~\cite{Hopfield1984,shiino1990stochastic,morita1993associative,inoue1996retrieval,Qiao2001}).
All of these studies assumed point-neurons, neural networks of arborized
neurons with non-additive coupling have not been studied in comparable
setups. Hopfield networks are paradigmatic models for associative
memory which may in particular contribute to solving two important
conundrums in Neuroscience: how biological neural networks achieve
a high memory capacity and how they can work so reliably under the
experimentally found noisy conditions. The incorporation of non-additive
dendrites into these models may therefore (1) shed light on the impact
of these features on memory capacity and robustness and, at the same
time, (2) allow to understand the underlying mechanisms due to their
analytical tractability.

In the first part of this article, we describe the response properties
of single neurons in presence of biologically plausible dendritic
nonlinearities in a statistical framework. In the second part, we
employ the results and study the effect of nonlinear dendrites on
associative memory networks. We consider networks of the Hopfield
type, as this standard model for associative memory lends itself to
analytical treatment and allows to concisely work out the effects
of nonlinear dendritic enhancement. We find that dendritic nonlinearities
improve pattern retrieval by effectively reducing the thresholds of
neurons and by increasing the robustness to noise. The improvement
is strongest for intermediate numbers of dendritic branches. We quantify
these effects and illustrate our analytical findings with numerical
simulations.

\section{Results}

\subsection{Basic model for non-additive processing in dendrites}

\begin{figure}[!t]
\noindent \begin{centering}
\includegraphics[width=1\columnwidth]{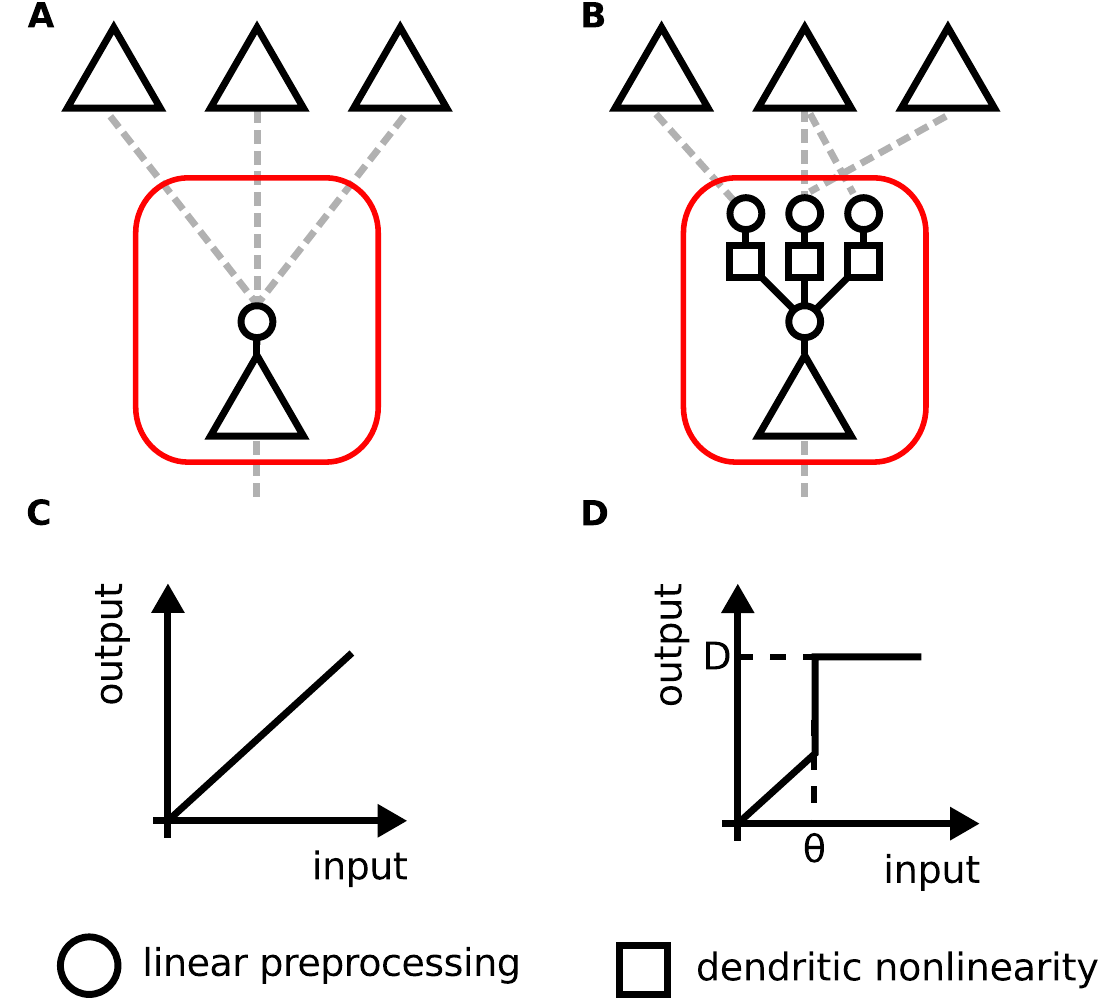}
\par\end{centering}

\protect\caption{\label{fig:model}\textbf{Arborized neuron with dendritic nonlinearities.}
Network architecture for a classical point-neuron model without dendrites
(\textbf{A}) and an arborized neuron with $B=3$ dendrites (or dendritic
branches) modeled as separate compartments (\textbf{B}). The red boxes
mark the neuronal unit for each case. Circles represent linear summation
of inputs (\textbf{C}), triangles and squares represent somatic and
dendritic processing, respectively. In a point-neuron all inputs are
summed up linearly and then processed nonlinearly. In a model with
nonlinear dendrites there is an additional, preceding layer where
inputs to each dendrite are summed up linearly and then subjected
to a dendritic nonlinearity. This nonlinearity is modeled as a piecewise
linear function with threshold $\theta$ and saturation strength $D$,
incorporating the effect of dendritic spikes (\textbf{D}). The somatic
transfer function is not constrained in the model.}
\end{figure}

Consider an extended topological structure of a neuron consisting
of one point-like soma and $B$ independent dendritic compartments
(Fig.~\ref{fig:model}). Each compartment receives its inputs from
a number of presynaptic neurons and transfers its output to the soma.
We assume that nonlinear dendritic integration takes place over a
time window $\Delta t$. Our model can be applied to dendrites with
fast or slow dendritic spikes, where $\Delta t$ assumes values of
$2-3\,\mathrm{ms}$ (fast, sodium spikes \cite{Ariav2003,Gasparini2004})
or tens of $\mathrm{ms}$ (slow, e.g.~NMDA spikes \cite{Schiller2000,Polsky2004}).
The durations of the different dendritic spikes have timescales similar
to their integration windows. The input arriving at a branch within
$\Delta t$ is denoted by $u$. On each branch, we capture the non-additive
input summation of the dendrites through a piecewise linear, sigmoidal
transfer function
\begin{eqnarray}
f\left(u\right) & := & \begin{cases}
u & \mbox{ if }u<\theta\\
D & \mbox{ otherwise}.
\end{cases}\label{eq:fu}
\end{eqnarray}
If$u$ is smaller than a threshold $\theta$, i.e.~$u<\theta$, the
inputs superpose linearly. Biologically, this means that $u$ has
not reached the threshold $\theta$ for dendritic spike generation
and that it is conventionally transferred to the soma. If the threshold
is exceeded, i.e.~$u\geq\theta$, the inputs superpose non-additively
and a fixed dendritic output strength $D$ is attained. This models
the effect of a dendritic spike elicited by sufficiently strong input.
The summation scheme as well as the compartmentalization are in agreement
with experimental findings and modeling studies \cite{Ariav2003,Poirazi2003,Gasparini2004,Polsky2004,Morita2007,Memmesheimer2010}.
Our approaches may be directly extended to neurons with multiple stages
of dendritic processing (cf.~App.~\ref{sec:Effective-input-approximation}).

\subsection{Capturing dendritic spikes by an effective somatic input strength}

To quantify the impact of non-additive dendritic events on the neuronal
input processing, the temporal and spatial distribution of synaptic
input must be taken into account. We consider a neuron with $B$ dendritic
branches $b\in\left\{ 1,\ldots,B\right\} $ and some time interval
of the length of the dendritic integration window. $x_{b}$ denotes
the number of synapses on branch $b$ that are active within this
window. The numbers of active synapses are distributed according to
$P\left(x_{1},\dots,x_{B}\right)$. Furthermore, we allow for distributed
connection strengths by assigning the synapses weights $w$ which
are independently and identically distributed according to $P\left(w\right)$.
Averages with respect to $P\left(x_{1},\dots,x_{B}\right)$ and $P\left(w\right)$
can be interpreted as ensemble averages or temporal averages. The
ensemble average is taken over a large number of neurons at a fixed
time where each neuron has numbers $x_{b}$ of active inputs and weights
$w$ which are samples of $P\left(x_{1},\dots,x_{B}\right)$ and $P\left(w\right)$,
respectively. Under the additional assumption of a large number of
synaptic contacts on each branch, the averages may also be understood
as time averages which are taken at a fixed neuron over a suitably
segmented long time interval in which the active inputs are changing.
In this article, we follow the first interpretation of ensemble averages.

What is the effective input to the soma given that synaptic inputs
are distributed across branches? We assume that each branch samples
a volume in which synapses of axons from $S$ other (presynaptic)
neurons can be synaptically contacted \cite{abeles1991corticonics}.
A synapse is present and active with probability $p_{b}$ such that
$P\left(x_{1},\dots,x_{B}\right)$ is a product of binomial distributions
with means $\mathrm{E}\left[x_{b}\right]=Sp_{b}$ and variances $\mathrm{Var}\left[x_{b}\right]=Sp_{b}\left(1-p_{b}\right)$.
Alternatively, the total number of active synaptic terminals across
branches might be fixed to $S$ (e.g.~due to homeostatic learning)
which suggests a multinomial distribution for $P\left(x_{1},\dots,x_{B}\right)$.
Then, the $x_{b}$ on different branches are not independent but negatively
correlated with covariances $\mathrm{Cov}\left[x_{b},x_{c}\right]=-Sp_{b}p_{c}$
for $b,c\in\left\{ 1,\dots,B\right\} $ and $b\neq c$.

The input to branch $b$ is given by the linear sum
\begin{eqnarray}
u_{b} & = & \sum_{i=1}^{x_{b}}w_{i}\label{eq:U}
\end{eqnarray}
and we are interested in the distribution $P\left(u_{1},\dots,u_{B}\right)$
of input across branches. According to Wald's equation \cite{wald1944cumulative},
the Blackwell-Girshick equation \cite{blackwell1946functions} and
the conditional covariance formula \cite{sheldon2002first}, we have
for $b\neq c$,

\begin{eqnarray}
\mathrm{E}\left[u_{b}\right] & = & \mathrm{E}\left[x_{b}\right]\mathrm{E}\left[w\right],\label{eq:meanU}\\
\mathrm{Var}\left[u_{b}\right] & = & \mathrm{E}\left[x_{b}\right]\mathrm{Var}\left[w\right]+\mathrm{Var}\left[x_{b}\right]\mathrm{E}^{2}\left[w\right],\label{eq:varU}\\
\mathrm{Cov}\left[u_{b},u_{c}\right] & = & \mathrm{E}\left[\mathrm{Cov}\left[u_{b},u_{c}\mid x_{b},x_{c}\right]\right]\nonumber \\
 &  & +\mathrm{Cov}\left[\mathrm{E}\left[u_{b}\mid x_{b},x_{c}\right],\mathrm{E}\left[u_{c}\mid x_{b},x_{c}\right]\right]\nonumber \\
 & = & \mathrm{Cov}\left[x_{b},x_{c}\right]\mathrm{E}^{2}\left[w\right],\label{eq:covU}
\end{eqnarray}
where $\mathrm{Cov}\left[u_{b},u_{c}\mid x_{b},x_{c}\right]=0$ and
$\mathrm{E}\left[u_{b}\mid x_{b}\right]=x_{b}\mathrm{E}\left[w\right]$
due to the independence of the synaptic weights $w$. $P\left(u_{1},\dots,u_{B}\right)$
is in general a complicated distribution (App.~\ref{sec:feff_exact}).
Since its first moments are known (Eqs.~\eqref{eq:meanU}-\eqref{eq:covU})
we approximate $P\left(u_{1},\dots,u_{B}\right)$ by a multivariate
normal distribution, i.e.~by the maximum entropy distribution for
the given moments.

This enables us to derive an effective input to the soma that depends
only on the number $B$ of branches, the probabilities $p_{b}$, $S$,
and the moments $\mathrm{E}\left[w\right]$ and $\mathrm{Var}\left[w\right]$.
For only linear branches, i.e.~$u_{b}<\theta$ on all branches, the
input to the soma is simply given by the linear sum $\sum_{b=1}^{B}u_{b}$.
For $u_{b}\geq\theta$, the dendritic nonlinearity sets in and branch
$b$ provides input of strength $D$ to the soma. Since somatic preprocessing
is linear, the total input to the soma is
\begin{eqnarray}
F & =F\left(u_{1},\ldots,u_{B}\right)= & \sum_{b=1}^{B}f\left(u_{b}\right).\label{eq:F}
\end{eqnarray}
The evaluation of the sum is numerically simple but denies analytical
treatment. Yet, for Gaussian $P\left(u_{1},\dots,u_{B}\right)$, we
may compute its mean (App.~\ref{sec:feff_gauss}) via
\begin{eqnarray}
\mathrm{E}\left[F\right] & = & \prod_{b=1}^{B}\int_{-\infty}^{\infty}\mathrm{d}u_{b}\left(\sum_{b=1}^{B}f\left(u_{b}\right)\right)P\left(u_{1},\dots,u_{B}\right)\nonumber \\
 & = & B\left(\int_{\theta}^{\infty}\mathrm{d}uDP\left(u\right)+\int_{-\infty}^{\theta}\mathrm{d}uuP\left(u\right)\right)\nonumber \\
 & = & BP_{\mathrm{NL}}D+B\left(1-P_{\mathrm{NL}}\right)E\left[u\right]-BC_{\mathrm{NL}},\label{eq:meanF}
\end{eqnarray}
where we exploited that the marginal distribution $P\left(u\right)$
of the multivariate normal distribution $P\left(u_{1},\dots,u_{B}\right)$
is a normal distribution again and defined
\begin{eqnarray}
P_{\mathrm{NL}} & := & \frac{1}{2}\mathrm{erfc}\left(\frac{\theta-\mathrm{E}\left[u\right]}{\sqrt{2\mathrm{Var}\left[u\right]}}\right),\label{eq:pnl}\\
C_{\mathrm{NL}} & := & \sqrt{\frac{\mathrm{Var}\left[u\right]}{2\pi}}\exp\left(-\frac{\left(\theta-\mathrm{E}\left[u\right]\right)^{2}}{2\mathrm{Var}\left[u\right]}\right).\label{eq:cnl}
\end{eqnarray}
In the second line of Eq.~\eqref{eq:meanF} we assumed $p_{b}=p_{0}$,
where $p_{0}$ is a constant probability independent of $b$. Throughout
the rest of the article we follow this choice for simplicity, although
many results are independent of $p_{b}$ or may be easily generalized
to arbitrary $p_{b}$. Eq.~\eqref{eq:meanF} may be interpreted as
follows: The first part describes the expected number $BP_{\mathrm{NL}}$
of dendritic spikes of strength $D$, while the second part captures
$B\left(1-P_{\mathrm{NL}}\right)$ linear events, and the last part
$BC_{\mathrm{NL}}$ corrects the overestimate of the contribution
of the linear branches by $\mathrm{E}\left[u\right]$. To obtain the
variance $\mathrm{Var}\left[F\right]=\mathrm{E}\left[F^{2}\right]-\mathrm{E}^{2}\left[F\right]$,
we analogously derive the second moment (App.~\ref{sec:feff_gauss})
\begin{eqnarray}
\mathrm{E}\left[F^{2}\right] & = & \prod_{b=1}^{B}\int_{-\infty}^{\infty}\mathrm{d}u_{b}\left(\sum_{b=1}^{B}f\left(u_{b}\right)\right)^{2}P\left(u_{1},\dots,u_{B}\right)\nonumber \\
 & = & B\int_{-\infty}^{\infty}\mathrm{d}uf^{2}\left(u\right)P\left(u\right)\nonumber \\
 &  & +\left(B^{2}-B\right)\int_{-\infty}^{\infty}\int_{-\infty}^{\infty}\mathrm{d}u\mathrm{d}vf\left(u\right)f\left(v\right)P\left(u,v\right).\label{eq:varF}
\end{eqnarray}
$P\left(u,v\right)$ is the marginal distribution of $P\left(u_{1},\dots,u_{B}\right)$
in two variables. For binomially distributed synapses, $P\left(u,v\right)=P\left(u\right)P\left(v\right)$
factors into two independent normal distributions which yields
\begin{eqnarray}
\mathrm{E}\left[F^{2}\right] & = & BP_{\mathrm{NL}}D^{2}+B\left(1-P_{\mathrm{NL}}\right)\left(\mathrm{E}^{2}\left[u\right]+\mathrm{Var}\left[u\right]\right)\nonumber \\
 &  & -BC_{\mathrm{NL}}\left(\mathrm{E}\left[u\right]+\theta\right)+\left(B^{2}-B\right)B^{-2}\mathrm{E}^{2}\left[F\right].\label{eq:varFbino}
\end{eqnarray}
For multinomially distributed synapses, the double integral in Eq.~\eqref{eq:varF}
needs to be evaluated numerically (see App.~\ref{sec:feff_gauss}
and Fig.~\ref{fig:feff}).

The mean $\mathrm{E}\left[F\right]$ and its variance $\mathrm{Var}\left[F\right]$
(as well as the expected number $\mathrm{E}\left[k\right]$ of nonlinear
branches and its variance $\mathrm{Var}\left[k\right]$, see below)
may also be computed without the Gaussian approximation employing
the exact expression for $P\left(u_{1},\dots,u_{B}\right)$ (App.~\ref{sec:feff_exact}).

We note that our approximation can be employed to compute the input
statistics to neurons with several layers of non-additive dendritic
branches by iteratively applying the formulas for $\mathrm{E}\left[F\right]$
and $\mathrm{Var}\left[F\right]$ (Eqs.~\eqref{eq:meanF}-\eqref{eq:varF})
to each branching point and using the result as a new input to the
next layer (cf.~App.~\ref{sec:Effective-input-approximation} for
a derivation).

\subsection{Features of the somatic input}

\begin{figure}[!t]
\noindent \begin{centering}
\includegraphics[width=1\columnwidth]{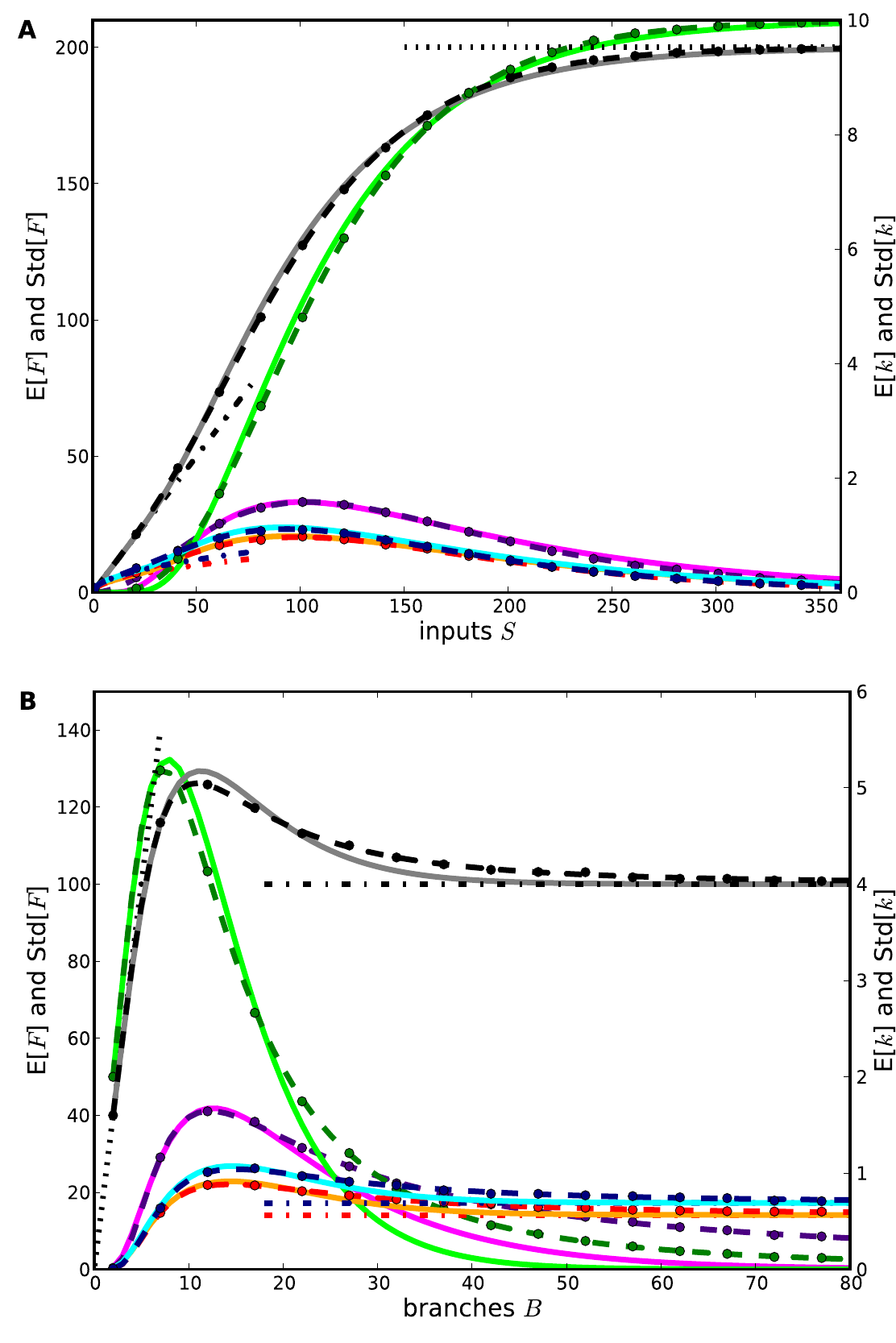}
\par\end{centering}

\protect\caption{\label{fig:feff}\textbf{Mean and standard deviation of the somatic
input $F$, optimal number of branches.} Results for fixed $B=10$
and varying $S$ (\textbf{A}) and vice versa, with $S=100$ (\textbf{B}).
Remaining parameters are $\theta=10$, $D=20$, $p_{0}=B^{-1}$ and
$P\left(w\right)$ is Gaussian with mean $\mathrm{E}\left[w\right]=1$
and variance $\mathrm{Var}\left[w\right]=2$. The effective somatic
input $\mathrm{E}\left[F\right]$, the average number of nonlinear
branches $\mathrm{E}\left[k\right]$ and their standard deviations
$\mathrm{Std}\left[F\right]=\sqrt{\mathrm{Var}\left[F\right]}$ and
$\mathrm{Std}\left[k\right]=\sqrt{\mathrm{Var}\left[k\right]}$ are
derived numerically using $2000$ realizations of binomially and multinomially
distributed synapses (circles). They agree well with the Gaussian
approximation (solid lines) and the exact analytical solution (dashed
lines). $\mathrm{E}\left[F\right]$ are shown in black (exact solution,
dashed) and gray (Gaussian approximation, solid), $\mathrm{Std}\left[F\right]$
are colored blue and cyan for binomially distributed synapses and
red and orange for multinomially distributed synapses. Averages $\mathrm{E}\left[k\right]$
are colored green and lime and their variances $\mathrm{Std}\left[k\right]$
for the binomial scenario are shown in purple and pink. Dash-dotted
and dotted lines represent the linear and saturated limits, respectively.
See main text for a detailed discussion.}
\end{figure}

We compare the input $F$ to the soma from numerical simulations (Eq.~\eqref{eq:F}
and Fig.~\ref{fig:feff}, circles), from the Gaussian approximation
(Eqs.~\eqref{eq:meanF} and\eqref{eq:varF} and Fig.~\ref{fig:feff},
solid lines), and the exact solution (App.~\ref{sec:feff_gauss}
and Fig.~\ref{fig:feff}, dashed lines) and find good agreement.
We choose $p_{0}=B^{-1}$ for both the binomial and the multinomial
case for a direct comparability of the two. Several features of the
input statistics may be noticed: The average $\mathrm{E}\left[F\right]$
is the same for the binomial and the multinomial distribution $P\left(x_{1},\dots,x_{B}\right)$
(Fig.~\ref{fig:feff}, Gaussian approximation in solid gray and exact
solution in dashed black) because their marginal distributions in
one variable are the same, in particular correlations among branches
do not contribute. The variation $\mathrm{Var}\left[F\right]$ is
larger in the binomial (Fig.~\ref{fig:feff}, solid cyan and dashed
blue) than in the multinomial scenario (Fig.~\ref{fig:feff}, solid
orange and dashed red) since the total number of active synaptic inputs
is constant in the latter and allows less fluctuation of input across
branches. In the strongly nonlinear regime, i.e.~$\mathrm{E}\left[u\right]\gg\theta$,
all branches are saturated and the average $\mathrm{E}\left[F\right]$
approaches saturation $DB$ (Fig.~\ref{fig:feff}, dotted black).
In the linear regime, i.e.~$\mathrm{E}\left[u\right]\ll\theta$,
Eq.~\eqref{eq:F} becomes $F=\sum_{i=1}^{x}w_{i}$, where $x=\sum_{b=1}^{B}x_{b}$
is the total number of active synapses. Then, the average $\mathrm{E}\left[F\right]=S\mathrm{E}\left[w\right]$
and hence grows linearly with the number of inputs $S$ but is independent
of $B$ (Fig.~\ref{fig:feff}, dash-dotted black). Further, the variance
$\mathrm{Var}\left[F\right]=S\mathrm{Var}\left[w\right]+S\left(1-B^{-1}\right)\mathrm{E}^{2}\left[w\right]$
in the binomial scenario (cf.~Eq.~\eqref{eq:varU}; Fig.~\ref{fig:feff},
dash-dotted blue) because $\mathrm{E}\left[x\right]=S$ and $\mathrm{Var}\left[x\right]=S\left(1-B^{-1}\right)$.
In the multinomial scenario, $\mathrm{Var}\left[F\right]=S\mathrm{Var}\left[w\right]$
(Fig.~\ref{fig:feff}, dash-dotted red) because the total number
of active synapses is fixed so that $\mathrm{Var}\left[x\right]=0$.

Another important feature of the mean somatic input $\mathrm{E}\left[F\right]$
is its maximum at an intermediate number $B=B_{F,\mathrm{opt}}$ of
branches (in Fig.~\ref{fig:feff}B, $B_{F,\mathrm{opt}}=11$). To
better understand this maximum, we compute the expectation value $\mathrm{E}\left[k\right]=BP_{\mathrm{NL}}$
of the number $k$ of branches in the nonlinear regime (and its variance
$\mathrm{Var}\left[k\right]=BP_{\mathrm{NL}}\left(1-P_{\mathrm{NL}}\right)$
in the binomial case; Gaussian approximation, cf.~App.~\ref{sec:feff_gauss}
and Fig.~\ref{fig:feff}, solid lime and pink; exact solution, cf.~App.~\ref{sec:feff_exact}
and Fig.~\ref{fig:feff}, dashed green and purple). $\mathrm{E}\left[k\right]$
has a maximum. This is plausible since the number of nonlinear branches
typically starts with one for $B=1$ because all input is concentrated
on this branch, then increases when more branches are available, but
goes to zero for large $B$. $\mathrm{E}\left[k\right]$ assumes its
maximum at approximately $B_{k,\mathrm{opt}}\approx\frac{S\mathrm{E}\left[w\right]}{\theta}$
because $\mathrm{Std}\left[u\right]\ll\mathrm{E}\left[u\right]$ for
(biologically plausible) sufficiently large ratios $SB^{-1}$ (Eqs.~\eqref{eq:meanU}
and \eqref{eq:varU}) so that $P_{\mathrm{NL}}$ (Eq.~\eqref{eq:pnl})
approaches a $\mathrm{step}$-function and $\mathrm{E}\left[k\right]=BP_{NL}\approx B\mathrm{step}\left(\theta-\frac{S}{B}\mathrm{E}\left[w\right]\right)$.
The somatic input $\mathrm{E}\left[F\right]$ (Eq.~\eqref{eq:meanF})
has two contributions, the first (from nonlinearly enhanced inputs)
is $D\mathrm{E}\left[k\right]$ and the second (from linearly summed
inputs) is monotonically increasing in $B$. Since $D$ is comparably
large, the maximum of $\mathrm{E}\left[k\right]$ induces a maximum
in $\mathrm{E}\left[F\right]$. The latter is shifted to the right
due to the monotonic increase of the linear contribution. The shift
indicates that a few additional branches may further increase $\mathrm{E}\left[F\right]$
because there synapses can provide input which would otherwise be
lost on saturated, nonlinear branches. Because a further increase
in the number of branches, however, leads to a substantial loss of
(non-additive) input, the maximum of $\mathrm{E}\left[F\right]$ is
close to that of $\mathrm{E}\left[k\right]$, i.e.~$B_{F,\mathrm{opt}}\approx\frac{S\mathrm{E}\left[w\right]}{\theta}$.

Up to now we considered combined processing of inhibition and excitation
on the dendritic branches. Often, inhibitory synapses are found to
directly target the soma \cite{Kim1995,gulyas1999total}. Such input
can be readily incorporated in our model by including an extra term
in Eq.~\eqref{eq:meanF},
\begin{eqnarray}
\mathrm{E}\left[F\right] & = & BP_{\mathrm{NL}}D+B\left(1-P_{\mathrm{NL}}\right)\mathrm{E}\left[u_{\mathrm{D}}\right]\nonumber \\
 &  & -BC_{\mathrm{NL}}+\mathrm{E}\left[u_{S}\right],\label{eq:meanFi}
\end{eqnarray}
where $\mathrm{E}\left[u_{D}\right]$ is the average input to a dendrite
and $P_{\mathrm{NL}}$ and $C_{\mathrm{NL}}$ (Eqs.~\eqref{eq:pnl}
and\eqref{eq:cnl}) are computed using $u=u_{D}$. $\mathrm{E}\left[u_{\mathrm{S}}\right]$
is the mean direct somatic input. Both summation scenarios may lead
to different collective dynamics on the network level (see below).

Concluding, we modeled the somatic input of a neuron with non-additive
dendrites. Our findings are independent of a specific neuron model.
We introduced a Gaussian approximation to describe the input irrespective
of the particular distribution of active synapses across branches
(Eqs.~\eqref{eq:meanF} and\eqref{eq:varF}). It provides a sufficiently
good description (cf.~Fig.~\ref{fig:feff}), simplifies calculations,
and is therefore used in the remainder of this article.

\subsection{Deterministic Hopfield networks of arborized neurons}

\begin{figure}[!t]
\noindent \begin{centering}
\includegraphics[width=1\columnwidth]{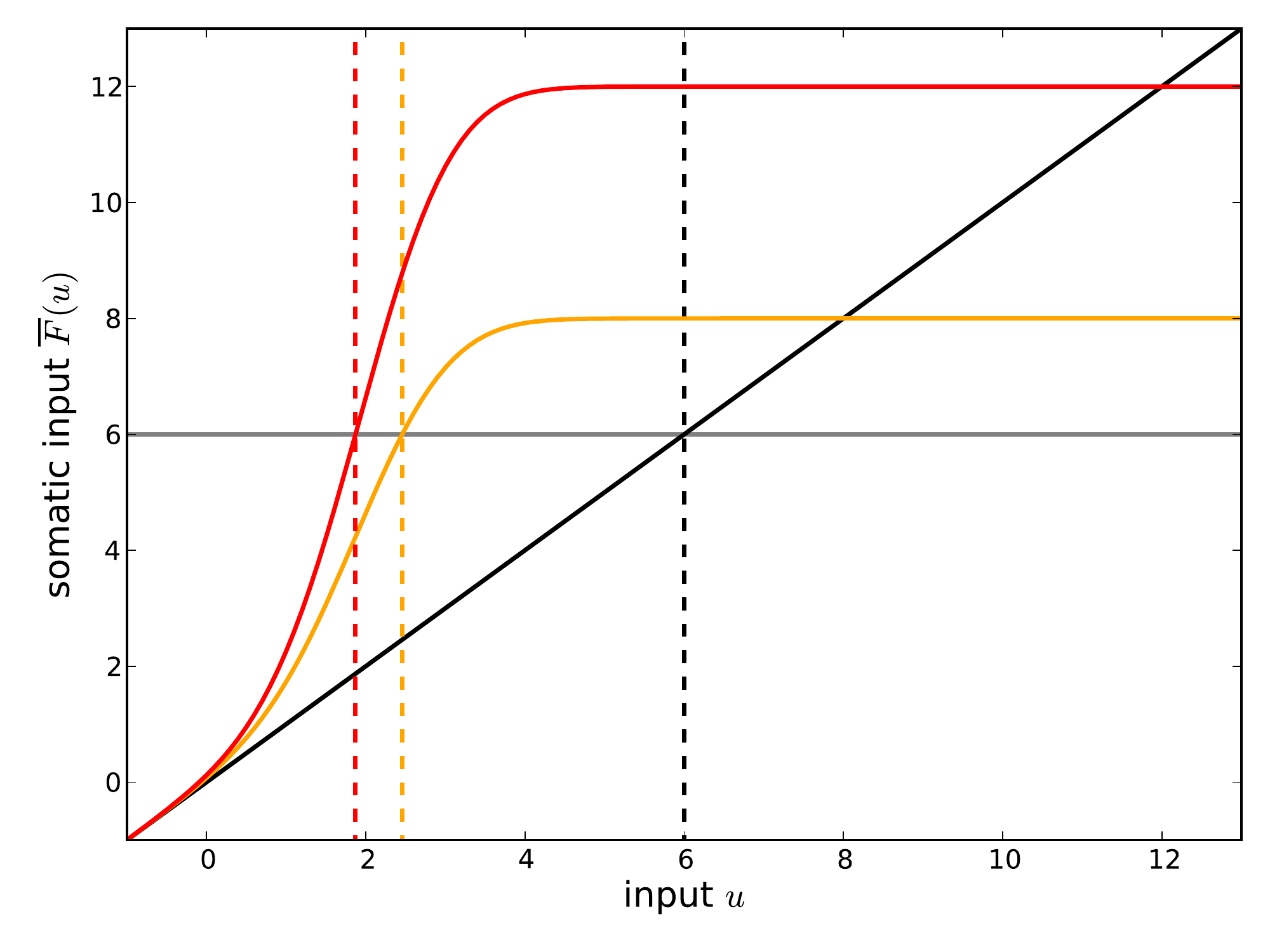}
\par\end{centering}

\protect\caption{\label{fig:thresred}\textbf{Effective reduction of the neuronal threshold
to $\vartheta\leq\Theta$ by nonlinear dendrites.} The traditional,
linear transfer function $u$ (solid black) crosses the threshold
$\Theta=6$ (solid gray) at $\vartheta=6$ (dashed black). For $B=2$,
$PN^{-1}\mathrm{Var}\left[w\right]=0.8$, and $\theta=1$, dendritic
nonlinearities $\bar{F}\left(u\right)$ (Eq.~\eqref{eq:F_explicitly_Hopfield})
with strengths $D=4$ (solid orange) and $D=6$ (solid red) lead to
effective thresholds $\vartheta\approx2.5$ (dashed orange) and $\vartheta\approx1.9$
(dashed red), respectively.}
\end{figure}

\begin{figure}[!t]
\noindent \begin{centering}
\includegraphics[width=1\columnwidth]{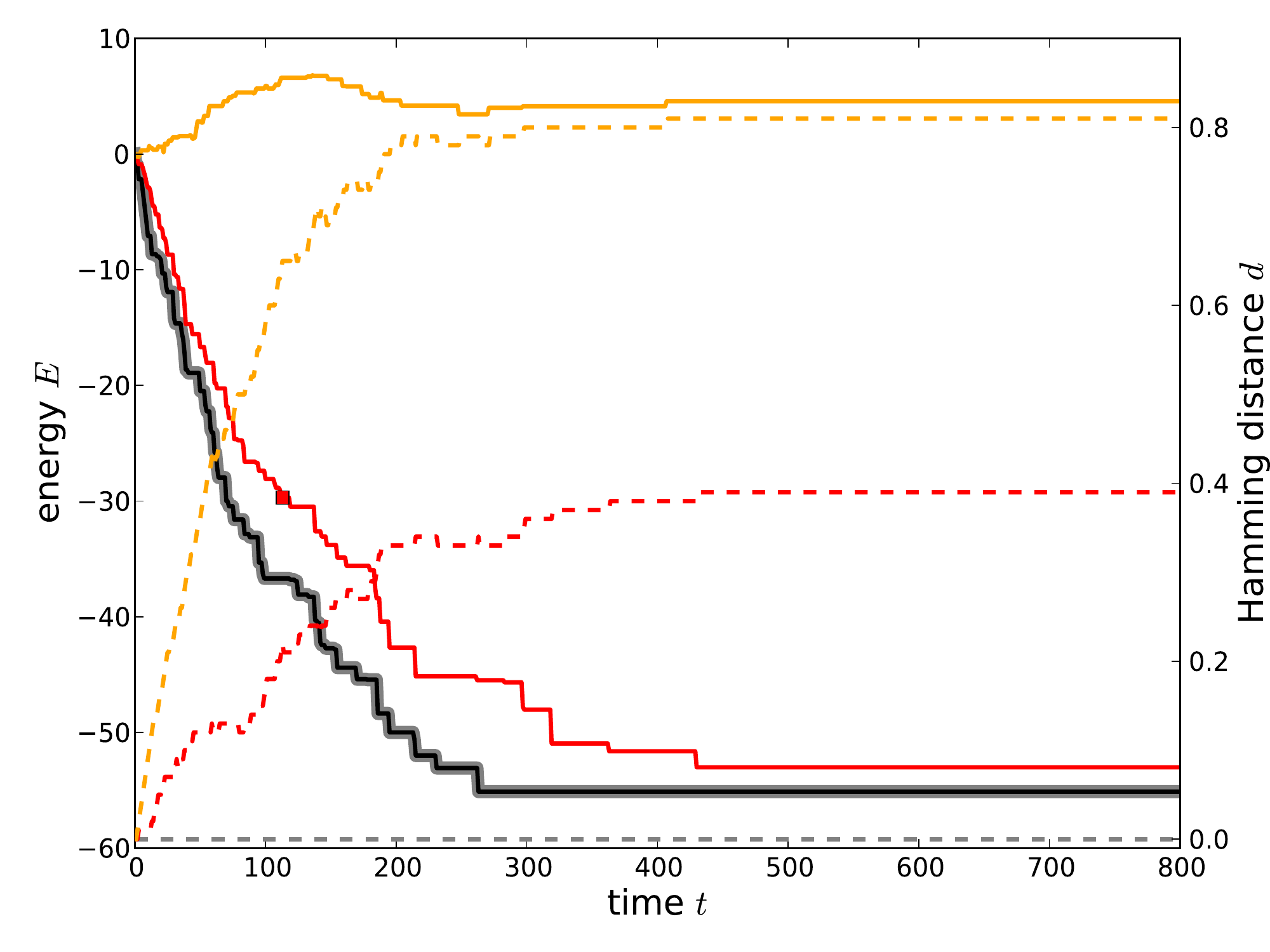}
\par\end{centering}

\protect\caption{\label{fig:conv}\textbf{Network convergence with effective threshold
reduction by nonlinear dendrites.} Simulation of a network of $N=100$
binary neurons with $\Theta=0.4$, $B=2$, $P=8$, and $\mathrm{Var}\left[w\right]=0.1$.
Simulations are performed with identical initial states, topology
and identical realization of the (random) order of updates. The energy
function (Eq.~\eqref{eq:Elin}) for linear input summation (solid
black) decreases and the network converges towards a fixed point.
Including weakly nonlinear branches (thick gray, $\theta=0.6$ and
$D=1$) does not alter the dynamics or the energy of the modified
network (which is now given by Eq.~\eqref{eq:Enonlin}). This is
confirmed by the vanishing Hamming distance $d=\frac{1}{2N}\sum_{n=1}^{N}\left|v_{n}-v_{n}^{'}\right|$
(dashed gray) between the two systems. For stronger dendritic spikes
(solid red, $\theta=0.1$ and $D=2$), deviations of the somatic input
from its ensemble average $\bar{F}$ and slightly asymmetric couplings
lead to occasional increases in the energy (red square) but network
convergence is preserved (checked for $1000$ runs). The dynamics
converge towards an attractor that is different from that of the linear
network as shown by the Hamming distance (dashed red) between the
systems. When inhibitory and excitatory inputs are processed separately
by the soma and the dendrites, respectively, no energy function is
known and we exemplarily choose the energy function given by Eq.~\eqref{eq:Enonlin}
with $\vartheta=\Theta$. The energy is then non-monotonic (solid
orange, $\theta=0.1$ and $D=2$) but the system nonetheless reaches
a stationary state (checked in simulations up to $t=2000$) which
is different from the one of the linear system, cf.~the Hamming distance
(dashed orange).}
\end{figure}

How do nonlinear dendrites influence the dynamics of associative neural
networks? Because of its analytical accessibility and its relevance
in neural computation, we consider a Hopfield network \cite{Hopfield1982}
of $N$ neurons $n\in\left\{ 1,\ldots,N\right\} $ with discrete states
$v_{n}\in\left\{ -1,+1\right\} $ and asynchronous updates of one
random unit at a discrete time $t\in\mathbb{N}$. We might interpret
$t$ as being measured in units of $N^{-1}\Delta t$ so that on average
each neuron is updated once per $\Delta t$ ($\approx$ dendritic
spike duration) and samples states which are present in other neurons
for $\Delta t$ ($\approx$ dendritic integration window). The update
rule for the conventional deterministic Hopfield model reads\textbf{
\begin{eqnarray}
v_{n}\left(t+1\right) & = & \mathrm{sign}\left[u_{n}\left(v_{1}\left(t\right),\dots,v_{N}\left(t\right)\right)-\Theta\right],\label{eq:update_det}
\end{eqnarray}
}where $n$ is the neuron updated at $t$. $\mathrm{sign}$$\left(x\right)$,
with $\mathrm{sign}\left(x\right)=-1$ if $x<0$ and $\mathrm{sign}\left(x\right)=1$
otherwise, is the neuronal transfer function and $\Theta$ denotes
the neuronal threshold. $u_{n}$ is the linear field
\begin{equation}
u_{n}=u_{n}\left(v_{1},\ldots,v_{N}\right)=\sum_{m=1}^{N}w_{n,m}v_{m}.\label{eq:field}
\end{equation}
The couplings $w_{n,m}$ between neurons $n$ and $m\in\left\{ 1,\ldots,N\right\} $
are assumed to be symmetric $w_{n,m}=w_{m,n}$. Then, a Lyapunov function
may be derived,\textbf{
\begin{eqnarray}
E_{\mathrm{L}}\left(v_{1}\left(t\right),\dots,v_{N}\left(t\right)\right) & = & -\frac{1}{2}\sum_{n=1}^{N}\sum_{m=1}^{N}w_{n,m}v_{n}\left(t\right)v_{m}\left(t\right)\nonumber \\
 &  & +\sum_{n=1}^{N}\Theta v_{n}\left(t\right),\label{eq:Elin}
\end{eqnarray}
}satisfying $E_{\mathrm{L}}\left(t+1\right)\leq E_{\mathrm{L}}\left(t\right)$
\cite{Hopfield1982,hertz1991introduction}. Equality $E_{\mathrm{L}}\left(t+1\right)=E_{\mathrm{L}}\left(t\right)$
only occurs if the state of the network upon update is not changed
or in the rare case when $u_{n}$ matches exactly the threshold $\Theta$.
Since $E_{\mathrm{L}}$ is bounded and $u_{n}=\Theta$ implies $v_{n}\left(t+1\right)=1$,
the weak Lyapunov property guarantees convergence of the system. Thus,
the network converges to an asymptotically stable minimum in the energy
landscape $E_{\mathrm{L}}\left(v_{1},\dots,v_{N}\right)$.

To store $P$ patterns $\xi^{p}$, $p\in\left\{ 1,\dots,P\right\} $,
the couplings in the Hopfield model are set in Hebbian manner \cite{hebb1949organization},
\begin{eqnarray}
w_{n,m} & = & N^{-1}\sum_{p=1}^{P}\xi_{n}^{p}\xi_{m}^{p}.\label{eq:hebb}
\end{eqnarray}
Classically, the storage of random, uncorrelated patterns $\xi_{n}^{p}\in\left\{ -1,+1\right\} $
is studied, where $\xi_{n}^{p}=\pm1$ with equal probabilities. Self-coupling
terms $w_{n,n}$ may lead to spurious states close to stored patterns
and are usually omitted, $w_{n,n}=0$\cite{hertz1991introduction}.
In this article we adopt these conventions. 

An alternative and similarly common model represents the neuronal
states via $v_{n}\in\left\{ 0,1\right\} $ (cf.~e.g.~\cite{tsodyks1988enhanced}).
For an appropriate choice of variables (cf.~\cite{hertz1991introduction})
this is equivalent to the $\left\{ -1,+1\right\} $-model, when introducing
a dependence of the neuronal threshold $\Theta$ on the couplings
$w_{nm}$. Since we assume constant Hebbian couplings (Eq.~\eqref{eq:hebb}),
we can include this term into the (then still constant) thresholds
and translate a $\left\{ 0,1\right\} $-model into a $\left\{ -1,+1\right\} $-model.
Because it is most often used in classical statistical mechanics studies
of neural networks (cf.~\cite{amit55storing}), we adopt the $\left\{ -1,+1\right\} $-representation.

We now modify this well-known model to incorporate dendritic branches.
For simplicity, we assume that each neuron has $B$ dendritic branches.
The arborization changes the network topology (Fig.~\ref{fig:model})
since neurons are now linked to branches. The coupling matrix becomes
a $N\times B\times N$-``matrix'' with entries $w_{n,b,m}$ that characterize
the coupling of neuron $m$ to branch $b$ of neuron $n$. The input
to an individual dendrite is given by the dendritic field
\begin{equation}
u_{n,b}=u_{n,b}\left(v_{1},\ldots,v_{N}\right)=\sum_{m=1}^{N}w_{n,b,m}v_{m}.
\end{equation}
The inputs are processed by the dendrites according to Eq.~\eqref{eq:fu}
and the somatic input is given by Eq.~\eqref{eq:F}. Taken together,
the update rule at time $t$ reads\textbf{
\begin{eqnarray}
v_{n}\left(t+1\right) & = & \mathrm{sign}\left[G_{n}\left(v_{1}\left(t\right),\dots,v_{N}\left(t\right)\right)-\Theta\right],\label{eq:update_nonlin_det}
\end{eqnarray}
}where
\begin{align}
G_{n} & =G_{n}\left(v_{1},\ldots,v_{N}\right)\nonumber \\
 & =F\left(u_{n,1}\left(v_{1},\ldots,v_{N}\right),\ldots,u_{n,B}\left(v_{1},\ldots,v_{N}\right)\right).\label{eq:Gn}
\end{align}

Like in the classical Hopfield model, we assume a Hebbian rule which
strengthens connections $w_{n,b,m}$ between co-active neurons such
that their expected value is $\mathrm{E}\left[w_{n,b,m}\right]=B^{-1}w_{n,m}$,
with $w_{n,m}$ given by Eq.~\eqref{eq:hebb}. Because the process
of adjustment of synaptic weights is subject to fluctuations \cite{graupner2012calcium},
we further assume that the weights $w_{n,b,m}$ are distributed with
variance $\mathrm{Var}\left[w_{n,b,m}\right]=w_{n,m}^{2}B^{-2}\mathrm{Var}\left[w\right]$.
The width of the distribution is proportional to the mean (with a
parameter $\mathrm{Var}\left[w\right]$) which avoids excessively
large deviations for small weights.

The network is fully connected and because input correlations across
branches vanish (App.~\ref{sec:distU}), this setup can be identified
with the binomial scenario introduced before, with $S=N$ and $p_{b}=p_{0}=1$.
Eqs.~\eqref{eq:meanU} and\eqref{eq:varU} yield (App.~\ref{sec:distU})
\begin{eqnarray}
\mathrm{E}\left[u_{n,b}\right] & = & B^{-1}u_{n},\label{eq:meanUn}\\
\mathrm{Var}\left[u_{n,b}\right] & = & B^{-2}PN^{-1}\mathrm{Var}\left[w\right].\label{eq:varUn}
\end{eqnarray}
The moments are computed as ensemble averages over an ensemble of
neurons with index $n$ at a fixed network state (annealed approximation).
The neural identity is preserved as it is specified by the parameters
$w_{n,m}$ that determine the expectation value and the variance of
the weight distribution of $w_{n,b,m}$ over which we average. An
averaging over $x_{b}$ is unnecessary as $p_{b}=p_{0}=1$, and thus
$\mathrm{E}\left[x_{b}\right]=N$ and $\mathrm{Var}\left[x_{b}\right]=0$.
The mean somatic input at neuron $n$, $\mathrm{E}\left[G_{n}\right]$,
follows from Eqs.~\eqref{eq:meanF}-\eqref{eq:cnl}. It depends on
$v_{1},\ldots,v_{N}$ and $n$ only via the mean input per branch
(Eq.~\eqref{eq:meanUn}), and thus via the linear field $u_{n}$
(Eq.~\eqref{eq:field}). We may therefore define an effective input
function $\bar{F}=\bar{F}\left(u_{n}\right)$ as
\begin{equation}
\bar{F}\left(u_{n}\left(v_{1},\ldots,v_{N}\right)\right)=\mathrm{E}\left[G_{n}\left(v_{1},\ldots,v_{N}\right)\right].\label{eq:G_and_F}
\end{equation}
From Eqs.~\eqref{eq:meanF}-\eqref{eq:cnl} we find
\begin{align}
\bar{F}\left(u_{n}\right) & =BP_{\mathrm{NL}}D+\left(1-P_{\mathrm{NL}}\right)u_{n}-BC_{\mathrm{NL}}\label{eq:F_explicitly_Hopfield}
\end{align}
with
\begin{eqnarray}
P_{\mathrm{NL}} & = & \frac{1}{2}\mathrm{erfc}\left(\frac{B\theta-u_{n}}{\sqrt{2PN^{-1}\mathrm{Var}\left[w\right]}}\right),\label{eq:pnl_explicitly_Hopfield}\\
C_{\mathrm{NL}} & = & B^{-1}\sqrt{\frac{P\mathrm{Var}\left[w\right]}{2\pi N}}\exp\left(-\frac{\left(B\theta-u_{n}\right)^{2}}{2PN^{-1}\mathrm{Var}\left[w\right]}\right).\,\,\,\,\,\,\,\label{eq:cnl_explicitly_Hopfield}
\end{eqnarray}
Analogously,Eq.~\eqref{eq:varFbino} shows that the standard deviation
of $G_{n}$, $\mathrm{Std}\left[G_{n}\right]$, is a function of the
mean input per branch only. We may therefore define $\mathrm{Std}\left[F\right]=\mathrm{Std}\left[F\left(u_{n}\right)\right]$
via $\mathrm{Std}\left[G_{n}\left(v_{1},\ldots,v_{n}\right)\right]=\mathrm{Std}\left[F\left(u_{n}\left(v_{1},\ldots,v_{n}\right)\right)\right]$
and compute it from Eqs.~\eqref{eq:varFbino} and\eqref{eq:F_explicitly_Hopfield}.

To investigate the convergence properties of the network, we consider
its state $\left(v_{1}\left(t\right),\dots,v_{N}\left(t\right)\right)$
at time $t$ and the update of a neuron $n$ by Eq.~\eqref{eq:update_nonlin_det}.
Led by Fig.~\ref{fig:feff}, we neglect the fluctuations $\mathrm{Std}\left[F\right]\ll\bar{F}$
and replace $G_{n}$ in Eq.~\eqref{eq:update_nonlin_det} by its
mean $\bar{F}\left(u_{n}\right)$. This approximation of the response
of neuron $n$ by the response of a typical neuron with identity $n$
(as specified by the weights $w_{n,m}$) simplifies the analysis of
the network dynamics. For reasonable parameter choices, the deviations
are small and lead to erroneous updates of neuron states very rarely
(Fig.~\ref{fig:conv}). Furthermore, we may assume that the effective
input function $\bar{F}\left(u_{n}\right)$ is strictly monotonic
in $u_{n}$ and thus invertible (App.~\ref{sec:conv}). These dynamics
are then equivalent to the conventional Hopfield network dynamics
(Eqs.~\eqref{eq:update_det}-\eqref{eq:Elin}, see App.~\ref{sec:conv})
with coupling matrix $w_{n,m}$ and an effective threshold $\vartheta$.
$\vartheta$ is determined by the intersection of the effective somatic
input $\bar{F}$ and the neuronal threshold $\Theta$ (Fig.~\ref{fig:thresred}),
\begin{equation}
\vartheta=\bar{F}^{-1}\left(\Theta\right).\label{eq:uc}
\end{equation}
In particular, for symmetric weights, $w_{n,m}=w_{m,n}$, the dynamics
has a Lyapunov function (App.~\ref{sec:conv})
\begin{eqnarray}
E_{\mathrm{NL}}\left(v_{1}\left(t\right),\dots,v_{N}\left(t\right)\right) & = & -\frac{1}{2}\sum_{n=1}^{N}\sum_{m=1}^{N}w_{n,m}v_{n}\left(t\right)v_{m}\left(t\right)\nonumber \\
 &  & +\sum_{n=1}^{N}\vartheta v_{n}\left(t\right).\label{eq:Enonlin}
\end{eqnarray}
By construction, $E_{\mathrm{NL}}$ decreases in time and the system
converges towards a dynamical fixed point. Thus, the supralinear dendrites
effectively reduce the neuronal threshold to $\vartheta\leq\Theta$
and leave the convergence properties of the system unchanged.

To study the convergence of the extended Hopfield model, we generate
a network by first drawing Hebbian synaptic weights $w_{n,m}$ according
to Eq.~\eqref{eq:hebb}. This yields random patterns with $\xi_{n}^{p}=\pm1$
with equal probabilities. Then, $w_{n,b,m}$ are drawn from a Gaussian
distribution with mean $\mathrm{E}\left[w_{n,b,m}\right]=B^{-1}w_{n,m}$
and variance $\mathrm{Var}\left[w_{n,b,m}\right]=w_{n,m}^{2}B^{-2}\mathrm{Var}\left[w\right]$
as explained above. We note that generally $\sum_{b=1}^{B}w_{n,b,m}=:w_{n,m}^{'}\neq w_{m,n}^{'}$
so that the neuronal connectivity in presence of dendrites is not
symmetric like in the classical Hopfield model. The energy function
in Eq.~\eqref{eq:Enonlin} was derived for symmetric weights $w_{n,m}^{'}=w_{n,m}=w_{m,n}=w_{m,n}^{'}$
which may be seen as an approximation valid at least for slightly
asymmetric couplings. Fig.~\ref{fig:conv} shows that this approximate
energy function correctly reflects the convergence of the network.
Also stronger deviations from the symmetric scenario (quantified by
$\mathrm{Var}\left[w\right]$) leave the findings largely unchanged
(App.~\ref{sec:asym}).

The results of numerical simulations displayed in Fig.~\ref{fig:conv}
illustrate the convergence properties of our model. If the threshold
reduction by dendrites is small, subthreshold inputs to a neuron remain
subthreshold also in the presence of nonlinear dendrites and the network
converges to the same state as for linear branches (Fig.~\ref{fig:conv},
gray). However, if the effective threshold reduction is stronger,
inputs that are subthreshold may become superthreshold due to the
dendritic nonlinearity and the same initial conditions tend to converge
towards different attractors (Fig.~\ref{fig:conv}, red). Since deviations
of the somatic input from its ensemble average $\bar{F}$ may violate
our approximation and the network is slightly asymmetric, the energy
function can occasionally increase (Fig.~\ref{fig:conv}, red square).
However, for the considered parameters, these events occur rarely
and do not affect the long-term convergence of the system (checked
for $1000$ runs, not shown).

If inhibitory synapses project directly onto the soma instead of being
mingled with excitatory synapses on the dendrites, the effective somatic
input $\bar{F}$ is given by Eq.~\eqref{eq:meanFi}. Since the dendritic
saturation by excitatory input may be exceeded by linear inhibition,
$\bar{F}$ is non-monotonic and no Lyapunov function is apparent.
The energy of the system as given by Eq.~\eqref{eq:Enonlin} with,
e.g., $\vartheta=\Theta$ does not decrease monotonically in time
(Fig.~\ref{fig:conv}, orange). However, numerical simulations suggest
that the network reaches a stable fixed point nevertheless, as exemplarily
shown in Fig.~\ref{fig:conv}. Such network convergence despite non-monotonic
transfer functions is known from other systems \cite{hirsch1989convergent,morita1993associative,inoue1996retrieval}.
These studies do not split excitation and inhibition but choose transfer
functions non-monotonic in the total, linear input $u_{n}$.

\subsection{Capacity of stochastic Hopfield networks with non-additive dendritic
input processing}

\begin{figure}[!t]
\noindent \begin{centering}
\includegraphics[width=1\columnwidth]{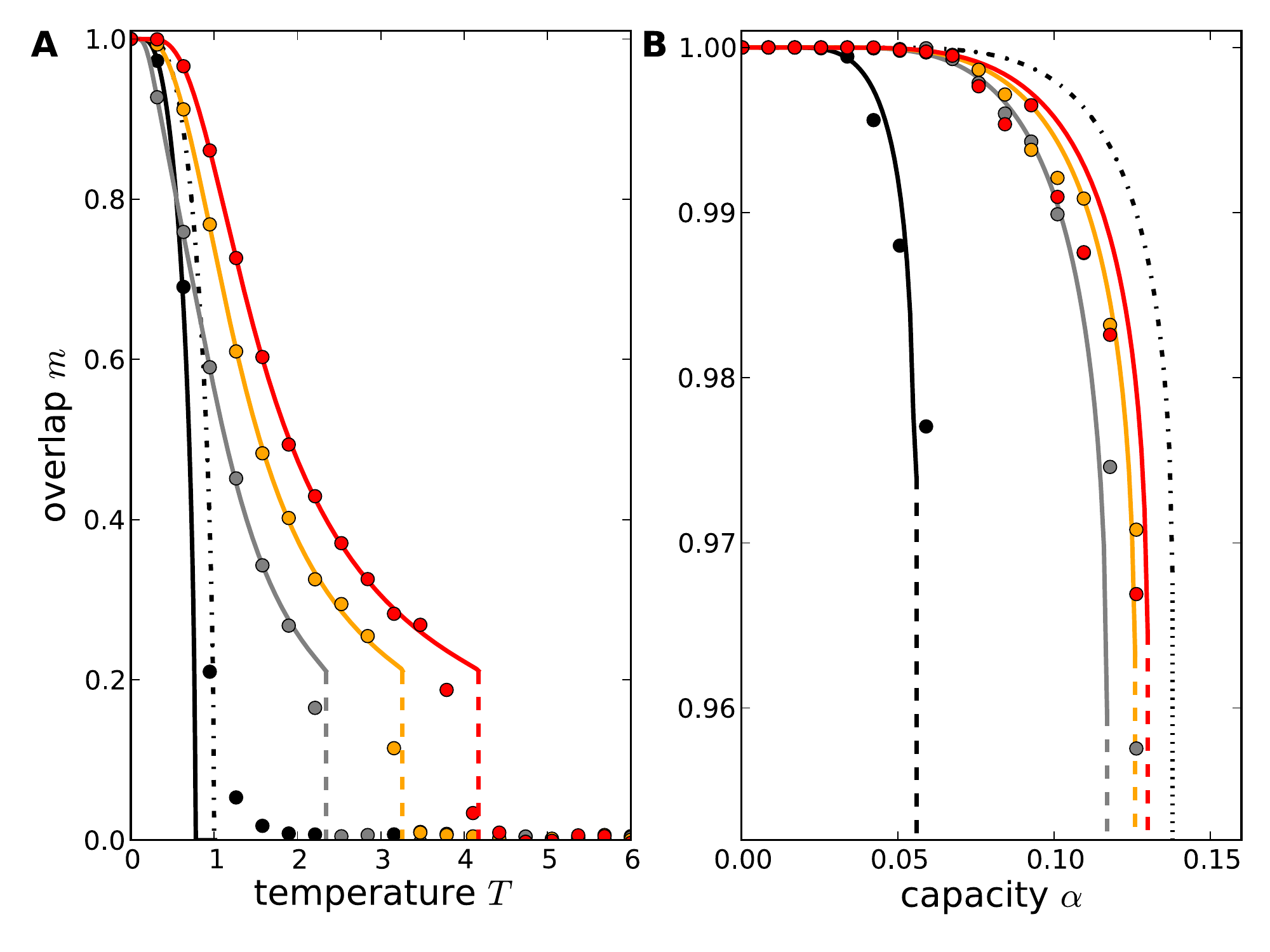}
\par\end{centering}

\protect\caption{\label{fig:mag}\textbf{Non-additive dendritic coupling increases
the overlap $m$ of the network state with a retrieval pattern.} Simulation
results (circles) and analytical results (solid lines) for linear
(black) and nonlinear summation with dendritic spike strengths $D\in\left\{ 0.4,0.6,0.8\right\} $
(gray, orange, red). The remaining parameters are $N=4000$, $B=2$,
$\theta=0.1$, $\Theta=0.4$, and $\mathrm{Var}\left[w\right]=0.1$.
Results for linear input summation with $\Theta=0$ are included for
comparison (dash-dotted black). The setup of the networks is the same
as described in the discussion of Fig.~\ref{fig:conv}. Panel (\textbf{A})
shows the overlap $m$ versus the temperature $T$ for a small load
$\alpha=N^{-1}\approx0$. It decreases with increasing $T$ and reaches
zero at the critical temperature $T_{c}$ above which retrieval fails.
This phase transition is discontinuous for the nonlinear and continuous
for the linear model of input processing. Dendritic nonlinearities
increase $T_{c}$. Panel (\textbf{B}) shows $m$ versus $\alpha$
at zero temperature $T=0$. It decreases with increasing load $\alpha$
and displays a discontinuous jump to zero at the critical storage
capacity $\alpha_{c}$. $\alpha_{c}$ increases with stronger dendritic
nonlinearities up to a level at which the effective threshold $\vartheta$
vanishes and the linear scenario with $\Theta=0$ is approached (dash-dotted
black).}
\end{figure}

\begin{figure}[!t]
\noindent \begin{centering}
\includegraphics[width=1\columnwidth]{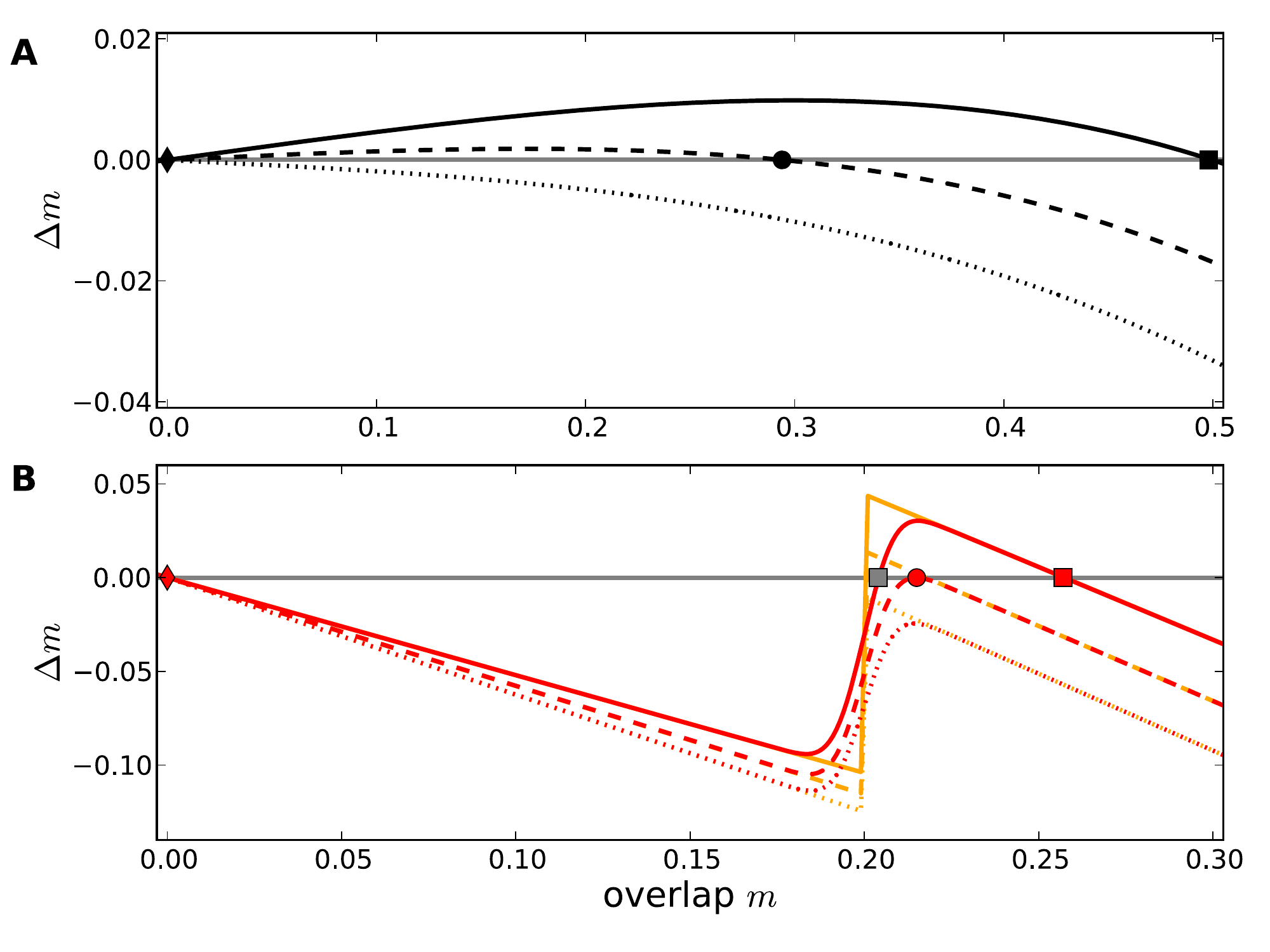}
\par\end{centering}

\protect\caption{\label{fig:phasetrans}\textbf{Graphical solutions to the transcendental
equation for the overlap $m$ and phase transitions.} Parameters are
the same as in Fig.~\ref{fig:mag}, with $D=0.4$. Eq.~\eqref{eq:mlima}
is solved graphically, with its solutions given by the zero-crossings
of $\Delta m$, i.e.~the intersections with the solid gray line.
For the traditional Hopfield case with linear input summation (\textbf{A}),
the stable solution for the overlap $m$ continuously decreases (black
markers, rectangle to circle to diamond) with increasing temperatures
$T\in\left\{ 0.7,0.75,0.8\right\} $ (solid, dashed, dotted black)
and reaches $m_{c}=0$ (diamond) at a critical value of $T_{c}\approx0.8$.
For nonlinear input summation due to nonlinear dendrites and small
$\alpha=N^{-1}\approx0$, panel (\textbf{B}) shows that increasing
temperatures $T\in\left\{ 2.0,2.295,2.6\right\} $ (solid, dashed,
dotted red) lead to decreasing overlaps $m$ (red markers, from rectangle
to circle to diamond) which jump discontinuously from $m_{c}\approx0.22$
(circle) to zero at $T_{c}\approx2.3$. The gray rectangle indicates
an unstable solution. In the limit $\alpha=0$ (B, orange), the jump
in $\Delta m$ is discontinuous, which slightly changes the critical
temperature but leaves the system's critical behavior unchanged.}
\end{figure}

We now assess the extent to which a network of binary neurons with
nonlinear dendrites is capable of storing and retrieving specific
patterns. Since biological neurons are noisy, i.e.~their input-output
relation is not fully reliable (e.g.~\cite{smetters1996synaptic}),
we generalize the above deterministic dynamics to allow for stochasticity.
For the analysis of the storage capacity of the extended Hopfield
network, we exploit the analogy between spin glasses and neural networks
and employ statistical physics methods \cite{Amit1985a,hertz1991introduction}.

As a generalization of the deterministic update rule (Eq.~\eqref{eq:update_nonlin_det})
we use the common Glauber dynamics \cite{glauber1963time,shiino1990stochastic,hertz1991introduction}
with asynchronous updates according to which the state of a randomly
chosen neuron $n$ is set to $v_{n}=\pm1$ with probability
\begin{eqnarray}
p_{n}\left(v_{n}\right) & = & \left(1+\exp\left(-2\beta\left[G_{n}-\Theta\right]v_{n}\right)\right)^{-1}.\label{eq:update_stoch}
\end{eqnarray}
This is equivalent to flipping the state of the respective neuron
with probability $p_{n}\left(v_{n}\rightarrow-v_{n}\right)=\left(1+\exp\left(2\beta\left[G_{n}-\Theta\right]v_{n}\right)\right)^{-1}$.
Here, $T:=\beta^{-1}$ is the pseudo-temperature and a measure for
the noise in the system and $G_{n}$ (Eq.~\eqref{eq:Gn}) is the
input to the neuron. We recall that $P$ is the number of desired
patterns and the fraction $\alpha=PN^{-1}$ is called load parameter.
We obtain a temporally averaged state $\left\langle v_{n}\right\rangle $
of unit $n$ in the ensemble-averaged, stochastic network in mean-field
theory by replacing $G_{n}$ by the ensemble average $\bar{F}\left(u_{n}\right)$
(Eq.~\eqref{eq:F_explicitly_Hopfield}). Further, we replace the
fluctuating argument of $\bar{F}\left(u_{n}\right)$ by its average,
which yields $\bar{F}\left(\left\langle u_{n}\right\rangle \right)$,
such that in the stationary state
\begin{eqnarray}
\left\langle v_{n}\right\rangle  & = & \left(+1\right)p_{n}\left(+1\right)+\left(-1\right)p_{n}\left(-1\right)\nonumber \\
 & = & \tanh\left(\beta\bar{F}\left(\left\langle u_{n}\right\rangle \right)-\beta\Theta\right)\nonumber \\
 & = & \tanh\left(\beta\bar{F}\left(N^{-1}\sum_{m=1}^{N}w_{n,m}\left\langle v_{m}\right\rangle \right)-\beta\Theta\right).\label{eq:meanfield}
\end{eqnarray}
The overlap between pattern $p$ and state $\left\langle v_{n}\right\rangle $
is defined by $m^{p}:=N^{-1}\sum_{n=1}^{N}\xi_{n}^{p}\left\langle v_{n}\right\rangle $.
Without loss of generality we study the retrieval of pattern $p=1$,
so that $m:=m^{1}$ estimates the quality of retrieval. $m$ is given
as an implicit solution to a set of coupled integral equations (App.~\ref{sec:meanfield},
Eqs.~\eqref{eq:app_Fm}, \eqref{eq:app_Fc}, \eqref{eq:app_Fr},
and\eqref{eq:app_Fs}). In particular, we consider two limits, $\alpha\approx0$
and $T=0$.

First, we study a finite number of patterns $P$ so that in the thermodynamic
limit of large $N$ we have $\alpha\approx0$ (App.~\ref{sec:lima}).
The overlap $m$ is given by the zeros of $\Delta m$,
\begin{eqnarray}
\Delta m & := & \frac{1}{2}\tanh\biggl(\beta(1-P_{\mathrm{NL}}\left(m\right))m\nonumber \\
 &  & +\beta\left(BDP_{\mathrm{NL}}\left(m\right)-BC_{\mathrm{NL}}\left(m\right)-\Theta\right)\biggr)\nonumber \\
 &  & +\frac{1}{2}\tanh\biggl(\beta(1-P_{\mathrm{NL}}\left(-m\right))m\nonumber \\
 &  & -\beta\left(BDP_{\mathrm{NL}}\left(-m\right)-BC_{\mathrm{NL}}\left(-m\right)-\Theta\right)\biggr)-m,\,\,\,\,\label{eq:mlima}
\end{eqnarray}
where the functions $P_{\mathrm{NL}}$ and $C_{\mathrm{NL}}$ are
given by Eqs.~\eqref{eq:pnl_explicitly_Hopfield} and\eqref{eq:cnl_explicitly_Hopfield}.
The solutions of the transcendental equation $\Delta m=0$ are obtained
numerically and compared to simulation results of the Hopfield network
with nonlinear dendrites (Fig.~\ref{fig:mag}A).

The dendritic nonlinearities have a strong impact on the overlap curve.
They change its shape and increase the critical temperature $T_{c}$
which marks the transition between functioning and non-functioning
associative memory. They provide a discontinuous, first order phase
transition with a non-zero critical overlap $m_{c}:=m\left(T_{c}\right)$.
For the same parameters, the conventional Hopfield model displays
a continuous, second order phase transition. These findings may be
understood by graphically solving Eq.~\eqref{eq:mlima}: For the
considered, not too large $\Theta$ ($\Theta<0.448$), linear input
processing leads to a concave $\Delta m\left(m\right)$ for high temperatures,
so that the overlap continuously goes to zero when $T$ approaches
the critical temperature $T_{c}$ (Fig.~\ref{fig:phasetrans}A).
In the presence of nonlinear dendrites, we need to take into account
that $PN^{-1}\mathrm{Var}\left[w\right]\propto\alpha\gtrsim0$ (Eqs.~\eqref{eq:F_explicitly_Hopfield}-\eqref{eq:cnl_explicitly_Hopfield})
is small so that the dendritic nonlinearity $\bar{F}$ sharply rises
to its maximal (saturation) value at $m$ with $B\theta-m\approx0$
due to its dependence on $P_{\mathrm{NL}}\left(m\right)$ (cf.~Eq.~\eqref{eq:pnl_explicitly_Hopfield}
with $u_{n}=m$; $C_{\mathrm{NL}}\left(m\right)\propto\alpha$ is
small). The second$\tanh$ is approximately constant there (because
$P_{\mathrm{NL}}\left(-m\right)$ and $C_{\mathrm{NL}}\left(-m\right)$
are small for small $PN^{-1}\mathrm{Var}\left[w\right]$) and may
be neglected. The sharp rise in $\bar{F}$ thus induces a convex turn
in the right-hand side of Eq.~\eqref{eq:mlima} and results in a
stable and an unstable fixed point of $m$ (Fig.~\ref{fig:phasetrans}B,
red). With growing temperature $T$, the two fixed points vanish in
a saddle-node bifurcation at non-zero $m_{c}\approx B\theta$ and
the system undergoes a discontinuous phase transition of first order.
For $\alpha=0$, the increase in $\bar{F}$ is jump-like so that no
unstable fixed point appears and the critical overlap is $m_{c}=B\theta$
(Fig.~\ref{fig:phasetrans}B, orange).

The increased critical temperature $T_{c}$ due to the dendrites implies
an increased robustness of the network against thermal fluctuations
and may be intuitively understood as follows: If the network state
is close to a learned pattern, the input to the neurons is either
strongly positive and thus further amplified by the dendritic nonlinearities
or strongly negative and not affected by the dendrites. The overall
strengthening of the stored patterns counteracts the influence of
the temperature (Eq.~\eqref{eq:update_stoch}) and stabilizes the
patterns against thermal fluctuations. Consequently, the nonlinear
dendrites allow pattern retrieval in a temperature regime in which
linear neurons fail.

Because for the conventional Hopfield model neuronal thresholds $\Theta$
decrease the critical temperature (cf.~Fig.~\ref{fig:mag}A, black),
we test if our results are a mere consequence of an effective threshold
reduction, $\vartheta\leq\Theta$, by the nonlinear dendrites (cf.~Eq.~\eqref{eq:uc}).
Repeating the above calculations and simulations with $\Theta=\vartheta=0$
we find that results as described above hold also for vanishing neuronal
thresholds (App.~\ref{sec:lima}).

Second, we consider the zero temperature limit $T=0$, in which thermal
fluctuations cease and the binary neurons are deterministic threshold
units (App.~\ref{sec:app_limb}). The overlap $m$ is determined
by
\begin{eqnarray}
m & = & \frac{1}{2}\mbox{erf}\left(\frac{m-\vartheta}{\sqrt{2\alpha r}}\right)+\frac{1}{2}\mbox{erf}\left(\frac{m+\vartheta}{\sqrt{2\alpha r}}\right),\nonumber \\
\sqrt{r} & = & 1+\sqrt{\frac{1}{2\pi\alpha}}\exp\left(-\frac{\left(m-\vartheta\right)^{2}}{2\alpha r}\right)\nonumber \\
 &  & +\sqrt{\frac{1}{2\pi\alpha}}\exp\left(-\frac{\left(m+\vartheta\right)^{2}}{2\alpha r}\right),\label{eq:mlimb}
\end{eqnarray}
where $\vartheta$ is the effective threshold (Eq.~\eqref{eq:uc}).
We solve these coupled equations numerically and compare them to the
simulation data of a Hopfield network with nonlinear dendrites (Fig.~\ref{fig:mag}B).
Eqs.~\eqref{eq:mlimb} are equivalent to the order parameter equations
of the conventional Hopfield model with threshold $\vartheta$. For
$\Theta>0$, we may therefore conclude that the dendritic branches
reduce the neuronal threshold to $\vartheta\leq\Theta$ and thereby
improve the critical storage capacity $\alpha_{c}$ of the network.
Analogous to $T_{c}$, $\alpha_{c}$ denotes the critical load above
which retrieval of patterns fails.

We note that the improved performance is a direct effect of the dendritic
nonlinearity as demonstrated in Fig.~\ref{fig:mag}, where the connectivity
of neurons is the same for networks with linear and nonlinear dendrites
while there is a clear increase in the critical temperature $T_{c}$
and load $\alpha_{c}$ in the latter case.

These findings may be understood by considering a neuronal threshold
$\Theta\neq0$ which generally introduces an asymmetry between the
two states $v_{n}=\pm1$ of a unit $n$. Since we assume the storage
of random patterns $\xi_{n}^{p}=\pm1$ with equal probabilities, the
non-zero $\Theta\neq0$ impedes the retrieval of learned patterns.
For a positive threshold $\Theta>0$, the threshold reduction by the
dendritic nonlinearities attenuates the asymmetry of the network and
improves the retrieval of random patterns. In the zero temperature
limit and for strong nonlinearities, our model becomes equivalent
to the standard Hopfield model without threshold $\Theta=0$ (Fig.~\ref{fig:mag}B,
dash-dotted black).

We note that the agreement of analytical and numerical results as
shown in Fig.~\ref{fig:mag} is even better if the couplings $w_{n,m}^{'}=\sum_{b=1}^{B}w_{n,b,m}$
are normalized such that the $w_{n,m}^{'}$ exactly equal the symmetric
Hebbian weights $w_{n,m}$. This holds in particular for stronger
dendritic spikes (Fig.~\ref{fig:mag}B, red) which emphasize the
asymmetry. To check if the above results hold also for larger deviations
from the assumption of symmetric couplings, $w_{n,m}^{'}=w_{m,n}^{'}$,
we repeat the simulations for larger $\mathrm{Var}\left[w\right]$
(cf.~Eq.~\eqref{eq:varUn}; App.~\ref{sec:asym}). In the deterministic
limit $T=0$ with many patterns, we find that stronger asymmetries
impede the quality of retrieval. For finite temperatures $T>0$ and
few patterns $\alpha\approx0$, the impact of moderately asymmetric
synaptic weights is negligible.

Complementing our analytical study of the limiting cases $\alpha\approx0$
and $T=0$, we compute the quality of retrieval in the $\alpha$-$T$-phase
space numerically. We find that our associative memory network with
non-additive dendrites enables memory functioning in a larger $\alpha$-$T$-region
than the model with linear branches (cf.~App.~\ref{sec:Comparison-of-phase}).

\subsection{Optimal number of dendritic branches for memory function}

\begin{figure}[!t]
\noindent \begin{centering}
\includegraphics[width=1\columnwidth]{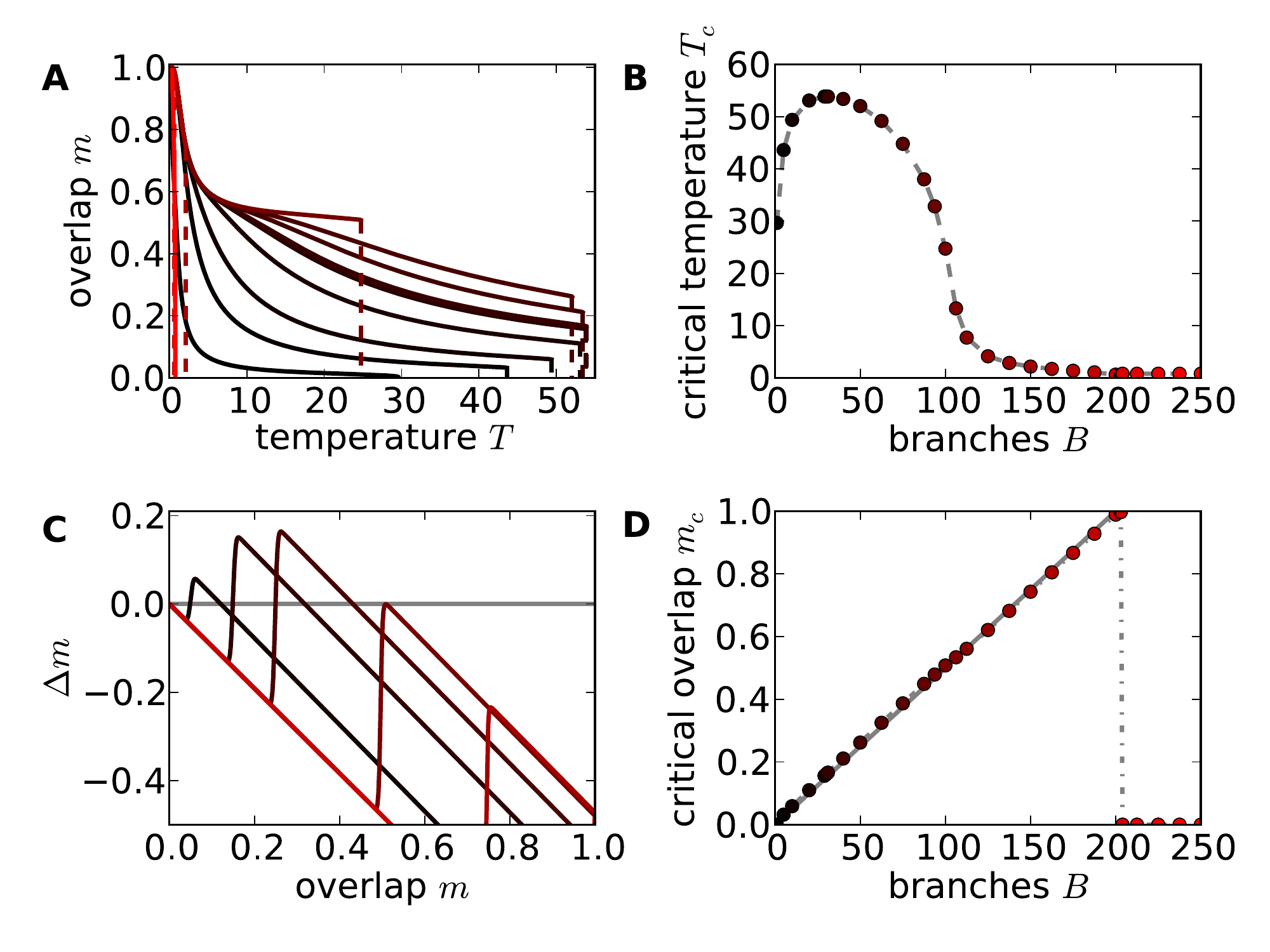}
\par\end{centering}

\protect\caption{\label{fig:bopt}\textbf{Memory performance in dependence of the number
$B$ of dendritic branches.} Parameters are the same as in Fig.~\ref{fig:mag},
with $D=0.6$ and $\theta=0.005$. Branch numbers $B$ are color-coded
from (black) to (red), cf.~panels (B) and (D). Panels (A) and (C)
show results for a few selected values of $B$. The overlap curves
$m\left(T\right)$ with $\alpha\approx0$ are computed for varying
numbers $B$ of branches (\textbf{A}). The critical temperature $T_{c}$
depends non-monotonically on $B$ and is maximal for $B_{T,\mathrm{opt}}=30$
(\textbf{B}, dashed line and circles). For a fixed temperature, $T=25$,
panel (\textbf{C}) shows $\Delta m$ whose zeros provide the solutions
for $m$ (Eq.~\eqref{eq:mlima}) and whose maxima display a non-monotonical
dependency on $B$. The critical overlap $m_{c}$ (\textbf{D}, dotted
line and circles) grows approximately linearly with $B$ as indicated
by the linear function $m_{c}=\theta B$ (\textbf{D}, solid line)
and jumps to zero at $m_{c}\approx1$.}
\end{figure}

Finally, we investigate the impact of varying numbers $B$ of dendritic
branches on the performance of the extended stochastic Hopfield network.
As shown above, non-additive dendritic input processing leads to an
increased storage capacity and more robust memory retrieval by amplifying
strong input to the neuron. Fig.~\ref{fig:feff}B shows that, when
this input is fixed, the average somatic input $\mathrm{E}\left[F\right]$
($\hat{=}\bar{F}$) is maximal for an intermediate number of dendrites.
This leads us to expect optimal memory performance for intermediate
branch numbers.

We thus study the Hopfield network for varying $B$. In analogy to
Fig.~\ref{fig:mag}A we compute the overlaps $m$ for the limiting
case $\alpha\approx0$ (Fig.~\ref{fig:bopt}A). The critical temperature
$T_{c}$ (Fig.~\ref{fig:bopt}B) displays a maximum at an intermediate
number of branches, here $B_{T,\mathrm{opt}}=30$. For larger $\theta$,
the optimal branch number is smaller (not shown). We can understand
the maximum in $T_{c}$ by considering the $\Delta m\left(m\right)$
curves (Fig.~\ref{fig:feff}C). They display maxima at $m_{c}\approx B\theta$
(see discussion of Fig.~\ref{fig:phasetrans}) with absolute heights
determined by $m_{c}$ and $\bar{F}\left(m_{c}\right)$ where the
latter is maximal for intermediate branch numbers. In combination,
they yield the maxima of $\Delta m\left(m\neq0\right)$ which are
highest for intermediate numbers $B_{T,\mathrm{opt}}$ of branches.
Upon increasing temperature, the corresponding curves are thus the
last to fall entirely below zero at their $T_{c}\left(B\right)$ so
that $T_{c}$ is highest for such branch numbers.

Another notable feature is the growing critical overlap $m_{c}$ at
the critical temperature $T_{c}$ with increasing numbers of branches
(Fig.~\ref{fig:bopt}D). As shown in the discussion of Fig.~\ref{fig:phasetrans},
the critical overlap for small $\alpha$ is given by $m_{c}\approx B\theta$.
The argument is correct for strong nonlinearities $D$ and moderate
$\theta B$ but breaks down for $\theta B\approx m_{c}>1$ since the
overlap is naturally bounded by $1$ and only the solution $m_{c}=0$
remains, if a larger overlap would be needed to reach the upturn point
of $\Delta m$ (cf.~Fig.~\ref{fig:bopt}C). In particular, for large
$B$, the behavior of the linear scenario with $m_{c}=0$ is reobtained.

Thus, the performance of the memory network depends non-trivially
on the number of dendritic branches $B$. Additionally, depending
on the purpose of the memory network, its robustness against noise
(specified by $T_{c}$) may be balanced against the quality of retrieval
(for which $m_{c}$ gives a worst-case measure) (cf.~Fig.~\ref{fig:bopt}A).

\section{Conclusion: Nonlinear dendrites improve pattern retrieval}

Non-additive processing of synaptic input is an important feature
of biological neurons and may have severe consequences for neural
processing in single neurons and networks. In this work, we first
studied the influence of dendritic spikes in a neuron with a variable
number of dendritic branches and sufficiently synchronous spiking
input of variable strength, independently of a specific neuron model.
We derived an approximation for the somatic input in the presence
of nonlinear dendrites (Eqs.~\eqref{eq:meanF}-\eqref{eq:varF} and
App.~\ref{sec:Effective-input-approximation}). This approximation
allows the analytical investigation of dendritic summation phenomena
in networks of arbitrary connectivity. Second, we extended the well-known
Hopfield model to include neurons with branches that process inputs
non-additively. Employing the results from the first part, we constructed
networks which are, at each neuron, ensemble-averaged over the nonlinear
dendrites and their inputs, such that the overall connectivity and
thus the neural identities in the networks are preserved. These networks
could be analyzed analytically with statistical physics methods. We
used them to approximate the full dynamics of networks with nonlinear
dendrites. We find that, for a deterministic Hopfield network, the
dendritic nonlinearities reduce the neuronal thresholds (Eq.~\eqref{eq:uc})
and the network still converges to a dynamical fixed point (Eq.~\eqref{eq:Enonlin}).
Separate processing of inhibition and excitation (Eq.~\eqref{eq:meanFi})
can break the monotonic decrease of common energy functions. A mean-field
analysis for a stochastic Hopfield network revealed an improved memory
storage capacity and a greater robustness against thermal fluctuations
due to non-additive dendritic input processing (Eqs.~\eqref{eq:mlima}
and \eqref{eq:mlimb} and App.~\ref{sec:Comparison-of-phase}). An
intermediate number of dendritic branches was shown to optimally support
memory functionality of the associative network.

Our findings help to advance the understanding of the role of nonlinear
dendrites in three respects: (i) Earlier works studied the ability
of arborized neurons to discriminate patterns \cite{Poirazi2001,Rhodes2008}.
They focused on a combinatorial approach of counting the numbers of
different input-output functions of single neurons with multiple dendrites.
In contrast, our study assumes a dynamical perspective. We derived
an expression for the approximate somatic input which may be readily
used to investigate the dynamics of networks of neurons with nonlinear
dendrites for arbitrary connectivity and synaptic weights. (ii) We
applied our results and studied the capability of networks with non-additive
dendrites to serve as memory devices. Our work shows that nonlinear
dendrites can increase the capacity and the robustness of memory retrieval
against thermal fluctuations in recurrent, dynamic associative memory
networks. (iii) Finally, our theoretical results suggest that there
might be an intermediate number of dendritic branches that is optimal
for network functionality \cite{Poirazi2001,Haeusser2003}. This may
have severe implications for biological neural circuits featuring
non-additive dendrites e.g.~in the hippocampus. Since non-additive
dendritic integration may take place in sliding window-like segments
of the dendritic tree, a precise number of independent dendritic compartments
is unlikely to be found \cite{Polsky2004}. Biological studies suggest
around $50-100$ independent sites capable of generating dendritic
spikes per neuron, depending on the neuron type and function \cite{Elston1998,Schiller2001}.
Our model shows optimal memory performance for such numbers of branches,
depending on the dendritic parameters (see discussion of Fig.~\ref{fig:bopt}).

For biological neural networks, we suggest that nonlinear dendrites
may serve to stabilize memory recall against noisy network background
activity. In such networks, memories might be stored in so-called
Hebbian cell assemblies with higher internal connectivities or connection
strengths that display elevated firing rates when presented with a
specific input pattern \cite{Sommer2001,Aviel2005}. Similar to our
Hopfield model, matching patterns of activity provide a larger input
to the other cells of the assembly which is amplified by the dendritic
nonlinearities. In our model we restricted ourselves to Hebbian learning
of coupling strengths, and noise and patterns were equally nonlinearly
enhanced. Biological neural networks can change their wiring as well
as the dendritic nonlinearities in an activity dependent manner \cite{Golding2002,Polsky2004,Losonczy2008}.
Further, some kinds of dendritic spikes amplify only temporally highly
coordinated inputs \cite{Ariav2003}. Both features may contribute
to a selective nonlinear enhancement of pattern activity and may increase
the stabilizing effects of non-additive dendrites in memory networks.

On a theoretical level of statistical physics, our work may be continued
in several directions: Preliminary calculations suggest that for certain
parameters new phenomena arise, such as improved memory retrieval
for moderate noise levels (this is reminiscent of stochastic resonance
\cite{Douglass1993,Kadar1998}). Further, previous studies on non-monotonic
transfer functions found beneficial effects for memory performance
in artificial and biological neural systems \cite{Hopfield1982,morita1993associative,inoue1996retrieval,Crespi1999}.
The novel kind of non-monotonicity which is due to dendritic reception
of excitatory input and somatic reception of inhibitory input (cf.~Eq.~\eqref{eq:meanFi})
should be further explored and linked to these findings. Finally,
many studies suggest that neural plasticity exploits dendritic spikes
and dendritic compartmentalization \cite{Golding2002,Polsky2004,Losonczy2008}.
Non-Hebbian learning rules that are tailored to utilize dendritic
spikes and branches were shown to increase the memory capabilities
of single neurons or ensembles of such neurons \cite{Engel1992,mel1992nmda,Poirazi2001,Rhodes2008,legenstein2011branch}.
Therefore, such dendrite-based learning is expected to boost also
the performance of associative memory networks and is a particularly
important target of future studies. Our statistical treatment (Eqs.~\eqref{eq:meanF}-\eqref{eq:varF})
and the numerical approach can be directly applied to address these
issues. Gaining insight into these matters will help to better understand
and utilize the full power of dendritic computation.

\section{Acknowledgments}

Supported by the BMBF (grant no.~01GQ1005B) and the DFG (grant no.~TI
629/3-1). We thank Sven Jahnke and Gunter Weber for helpful discussions.

\bibliographystyle{apalike}

\cleardoublepage{}

\cleardoublepage
\appendix
\onecolumngrid
\setcounter{page}{1}
\renewcommand{\thepage}{A\arabic{page}}
\setcounter{table}{0}
\renewcommand{\thetable}{A\arabic{table}}
\setcounter{figure}{0}
\renewcommand{\thefigure}{A\arabic{figure}}
\setcounter{section}{0}
\renewcommand{\thesection}{A\arabic{section}}
\makeatletter
\numberwithin{equation}{section}
\numberwithin{figure}{section}

\section{Approximate mean and variance of the effective somatic input and
the number of nonlinear branches\label{sec:feff_gauss}}

We compute the first moments of the somatic input $F$ (Eq.~\eqref{eq:F})
using a Gaussian approximation of the dendritic input distribution
$P\left(u_{1},\dots,u_{B}\right)$ and assuming statistically identical
branches, i.e.~$p_{b}=p_{0}$, $b\in\left\{ 1,\dots,B\right\} $.
Means $\mathrm{E}\left[u_{b}\right]$, variances $\mathrm{Var}\left[u_{b}\right]$
and covariances $\mathrm{Cov}\left[u_{b},u_{c}\right]$ of $P\left(u_{1},\dots,u_{B}\right)$
are given by Eqs.~\eqref{eq:meanU}-\eqref{eq:covU}. We start with

\begin{eqnarray}
\mathrm{E}\left[F\right] & = & \int_{-\infty}^{\infty}\mathrm{d}u_{1}\dots\int_{-\infty}^{\infty}\mathrm{d}u_{B}\left(f\left(u_{1}\right)+\dots+f\left(u_{B}\right)\right)P\left(u_{1},\dots,u_{B}\right)\label{eq:app_EF}
\end{eqnarray}
and pick $u:=u_{1}$ without loss of generality to obtain

\begin{eqnarray}
\mathrm{E}\left[F\right] & = & B\int_{-\infty}^{\infty}\mathrm{d}uf\left(u\right)\int_{-\infty}^{\infty}\mathrm{d}u_{2}\dots\int_{-\infty}^{\infty}\mathrm{d}u_{B}P\left(u,u_{2,}\dots,u_{B}\right)\nonumber \\
 & = & B\int_{-\infty}^{\infty}\mathrm{d}uf\left(u\right)P\left(u\right),
\end{eqnarray}
where $P\left(u\right)$ is the marginal distribution of $P\left(u,u_{2},\dots,u_{B}\right)$
and thus Gaussian with mean $\mathrm{E}\left[u\right]$ and variance
$\mathrm{Var}\left[u\right]$. Using the definition of the dendritic
transfer function $f$ (Eq.~\eqref{eq:fu}), we split

\begin{eqnarray}
\mathrm{E}\left[F\right] & = & B\left(\int_{\theta}^{\infty}\mathrm{d}uDP\left(u\right)+\int_{-\infty}^{\theta}\mathrm{d}uuP\left(u\right)\right)\nonumber \\
 & = & BP_{\mathrm{NL}}D+B\left(1-P_{\mathrm{NL}}\right)\mathrm{E}\left[u\right]-BC_{\mathrm{NL}},\label{eq:app_ef_approx}
\end{eqnarray}
where we used partial integration in the second line and the definitions
\begin{eqnarray}
P_{\mathrm{NL}} & := & \frac{1}{2}\mathrm{erfc}\left(\frac{\theta-\mathrm{E}\left[u\right]}{\sqrt{2\mathrm{Var}\left[u\right]}}\right),\\
C_{\mathrm{NL}} & := & \sqrt{\frac{\mathrm{Var}\left[u\right]}{2\pi}}\exp\left(-\frac{\left(\theta-\mathrm{E}\left[u\right]\right)^{2}}{2\mathrm{Var}\left[u\right]}\right).
\end{eqnarray}
The second moment is computed similarly,
\begin{eqnarray}
\mathrm{E}\left[F^{2}\right] & = & \int_{-\infty}^{\infty}\mathrm{d}u_{1}\dots\int_{-\infty}^{\infty}\mathrm{d}u_{B}\left(f\left(u_{1}\right)+\dots+f\left(u_{B}\right)\right)^{2}P\left(u_{1},\dots,u_{B}\right)\nonumber \\
 & = & B\int_{-\infty}^{\infty}\mathrm{d}uf^{2}\left(u\right)P\left(u\right)+\left(B^{2}-B\right)\int_{-\infty}^{\infty}\mathrm{d}u\int_{-\infty}^{\infty}\mathrm{d}vf\left(u\right)f\left(v\right)P\left(u,v\right)\nonumber \\
 & = & B\left(\int_{\theta}^{\infty}\mathrm{d}uD^{2}P\left(u\right)+\int_{-\infty}^{\theta}\mathrm{d}uu^{2}P\left(u\right)\right)\nonumber \\
 &  & +\left(B^{2}-B\right)\Biggl(\int_{\theta}^{\infty}\mathrm{d}u\int_{\theta}^{\infty}\mathrm{d}vD^{2}P\left(u,v\right)\nonumber \\
 &  & +2\int_{\theta}^{\infty}\mathrm{d}u\int_{-\infty}^{\theta}\mathrm{d}vDuP\left(u,v\right)+\int_{-\infty}^{\theta}\mathrm{d}u\int_{-\infty}^{\theta}\mathrm{d}vuvP\left(u,v\right)\Biggr)\nonumber \\
 & = & BP_{\mathrm{NL}}D^{2}+B\left(1-P_{\mathrm{NL}}\right)\left(\mathrm{E}^{2}\left[u\right]+\mathrm{Var}\left[u\right]\right)-BC_{\mathrm{NL}}\left(\mathrm{E}\left[u\right]+\theta\right)\nonumber \\
 &  & +\left(B^{2}-B\right)I_{F}.
\end{eqnarray}
Here, $v:=u_{2}$ and the marginal distribution $P\left(u,v\right)$
is again Gaussian with means $\left(\mathrm{E}\left[u\right],\mathrm{E}\left[v\right]\right)$,
variances $\left(\mathrm{Var}\left[u\right],\mathrm{Var}\left[v\right]\right)$
and covariance $\mathrm{Cov}\left[u,v\right]$. For binomially distributed
numbers of active synapses per branch, $P\left(u,v\right)=P\left(u\right)P\left(v\right)$
and therefore
\begin{eqnarray}
I_{F} & = & \left(P_{\mathrm{NL}}D\right)^{2}+2\left(P_{\mathrm{NL}}D\right)\left(\left(1-P_{\mathrm{NL}}\right)\mathrm{E}\left[u\right]-C_{\mathrm{NL}}\right)+\left(\left(1-P_{\mathrm{NL}}\right)\mathrm{E}\left[u\right]-C_{\mathrm{NL}}\right)^{2}\nonumber \\
 & = & \left(P_{\mathrm{NL}}D+\left(1-P_{\mathrm{NL}}\right)\mathrm{E}\left[u\right]-C_{\mathrm{NL}}\right)^{2}\nonumber \\
 & = & \left(B^{-1}\mathrm{E}\left[F\right]\right)^{2}.
\end{eqnarray}
For a multinomial distribution of active synapses across branches,
the double integral $I_{F}$ needs to be computed numerically (see
Fig.~\ref{fig:feff}).These results are discussed in the main text
and in the caption to Fig.~\ref{fig:feff}. The calculations may
be easily extended to cover non-uniform branch probabilities $p_{b}$,
dendritic thresholds $\theta_{b}$, and strengths $D_{b}$.

Similar to $\mathrm{E}\left[F\right]$, we compute the expected number
$\mathrm{E}\left[k\right]$ of branches $k$ in the nonlinear regime.
The dendritic transfer function $f\left(u\right)$ in Eq.~\eqref{eq:app_EF}
is replaced by a step function $\mathrm{step}\left(u-\theta\right)$
to count the number of branches above threshold $\theta$. Here, we
defined $\mathrm{step}\left(x\right)=0$ if $x<0$ and $\mathrm{step}\left(x\right)=1$
otherwise. Then,
\begin{eqnarray}
\mathrm{E}\left[k\right] & = & \int_{-\infty}^{\infty}\mathrm{d}u_{1}\dots\int_{-\infty}^{\infty}\mathrm{d}u_{B}\left(\mathrm{step}\left(u_{1}-\theta\right)+\dots+\mathrm{step}\left(u_{B}-\theta\right)\right)P\left(u_{1},\dots,u_{B}\right)\nonumber \\
 & = & B\int_{\theta}^{\infty}\mathrm{d}uP\left(u\right)\nonumber \\
 & = & BP_{\mathrm{NL}}.
\end{eqnarray}

Since the step function satisfies $\mathrm{step}^{2}\left(x\right)=\mathrm{step}\left(x\right)$
we derive the second moment of the distribution of the number $k$
of nonlinear branches via
\begin{eqnarray}
\mathrm{E}\left[k^{2}\right] & = & \int_{-\infty}^{\infty}\mathrm{d}u_{1}\dots\int_{-\infty}^{\infty}\mathrm{d}u_{B}\left(\mathrm{step}\left(u_{1}-\theta\right)+\dots+\mathrm{step}\left(u_{B}-\theta\right)\right)^{2}P\left(u_{1},\dots,u_{B}\right)\nonumber \\
 & = & B\int_{\theta}^{\infty}\mathrm{d}uP\left(u\right)+\left(B^{2}-B\right)\int_{\theta}^{\infty}\mathrm{d}u\int_{\theta}^{\infty}\mathrm{d}vP\left(u,v\right)\nonumber \\
 & = & BP_{\mathrm{NL}}+\left(B^{2}-B\right)I_{k}.
\end{eqnarray}
The binomial distribution of synapses among branches provides $P\left(u,v\right)=P\left(u\right)P\left(v\right)$
and thus
\begin{eqnarray}
I_{k} & = & P_{\mathrm{NL}}^{2}=\left(B^{-1}\mathrm{E}\left[k\right]\right)^{2},
\end{eqnarray}
while $I_{k}$ needs to be computed numerically in the multinomial
case.

\section{Exact mean and variance of the effective somatic input and the number
of nonlinear branches\label{sec:feff_exact}}

We now derive the mean somatic input $\mathrm{E}\left[F\right]$ and
its variance $\mathrm{Var}\left[F\right]$ for the exact distribution
$P\left(u_{1},\dots,u_{B}\right)$ of input where $u_{b}=\sum_{i=1}^{x_{b}}w_{i}$
(cf.~Eq.~\eqref{eq:U}) with Gaussian $P\left(w\right)$. We decompose
$P\left(u_{1},\dots,u_{B}\right)$ into
\begin{eqnarray}
P\left(u_{1},\dots,u_{B}\right) & = & \sum_{x_{1}=1}^{S}\dots\sum_{x_{B}=1}^{S}P\left(u_{1},\dots,u_{B}\mid x_{1},\dots,x_{B}\right)P\left(x_{1},\dots,x_{B}\right),
\end{eqnarray}
where
\begin{eqnarray}
P\left(u_{1},\dots,u_{B}\mid x_{1},\dots,x_{B}\right) & = & P\left(u_{1}\mid x_{1}\right)\cdots P\left(u_{B}\mid x_{B}\right)
\end{eqnarray}
since $u_{b}$ depends only on $x_{b}$. The $P\left(u_{b}\mid x_{b}\right)$
are Gaussian distributed with means $x_{b}\mathrm{E}\left[w\right]$
and variances $x_{b}\mathrm{Var}\left[w\right]$ and $P\left(u_{1},\dots,u_{B}\right)$
is a weighted superposition of Gaussian distributions. For non-Gaussian
$P\left(w\right)$ and large numbers of small inputs, we may employ
a central limit theorem to establish Gaussianity of the $P\left(u_{b}\mid x_{b}\right)$.
$\mathrm{E}\left[F\right]$ is computed as

\begin{eqnarray}
\mathrm{E}\left[F\right] & = & \int_{-\infty}^{\infty}\mathrm{d}u_{1}\dots\int_{-\infty}^{\infty}\mathrm{d}u_{B}\left(f\left(u_{1}\right)+\dots+f\left(u_{B}\right)\right)P\left(u_{1},\dots,u_{B}\right)\nonumber \\
 & = & \int_{-\infty}^{\infty}\mathrm{d}u_{1}\dots\int_{-\infty}^{\infty}\mathrm{d}u_{B}\sum_{b=1}^{B}f\left(u_{b}\right)\sum_{x_{1}=1}^{S}\dots\sum_{x_{B}=1}^{S}P\left(x_{1},\dots,x_{B}\right)P\left(u_{1},\dots,u_{B}\mid x_{1},\dots,x_{B}\right)\nonumber \\
 & = & \int_{-\infty}^{\infty}\mathrm{d}u_{1}\dots\int_{-\infty}^{\infty}\mathrm{d}u_{B}\sum_{b=1}^{B}f\left(u_{b}\right)\sum_{x_{1}=1}^{S}\dots\sum_{x_{B}=1}^{S}P\left(x_{1},\dots,x_{B}\right)P\left(u_{1}\mid x_{1}\right)\cdots P\left(u_{B}\mid x_{B}\right)\nonumber \\
 & = & \sum_{x_{1}=1}^{S}\dots\sum_{x_{B}=1}^{S}P\left(x_{1},\dots,x_{B}\right)\sum_{b=1}^{B}\int_{-\infty}^{\infty}\mathrm{d}u_{b}f\left(u_{b}\right)P\left(u_{b}\mid x_{b}\right)\prod_{c\neq b}^{B}\int_{-\infty}^{\infty}\mathrm{d}u_{c}P\left(u_{c}\mid x_{c}\right)\nonumber \\
 & = & \sum_{x_{1}=1}^{S}\dots\sum_{x_{B}=1}^{S}P\left(x_{1},\dots,x_{B}\right)\sum_{b=1}^{B}\int_{-\infty}^{\infty}\mathrm{d}u_{b}f\left(u_{b}\right)P\left(u_{b}\mid x_{b}\right)\cdot1\nonumber \\
 & = & \sum_{x_{1}=1}^{S}\dots\sum_{x_{B}=1}^{S}P\left(x_{1},\dots,x_{B}\right)\sum_{b=1}^{B}\left[P_{\mathrm{NL},u_{b}\mid x_{b}}D+\left(1-P_{\mathrm{NL},u_{b}\mid x_{b}}\right)\mathrm{E}\left[u_{b}\mid x_{b}\right]-C_{\mathrm{NL},u_{b}\mid x_{b}}\right]\nonumber \\
 & = & \sum_{x=1}^{S}P\left(x\right)B\left[P_{\mathrm{NL},u\mid x}D+\left(1-P_{\mathrm{NL},u\mid x}\right)\mathrm{E}\left[u\mid x\right]-C_{\mathrm{NL},u\mid x}\right],\label{eq:app_ef_exact}
\end{eqnarray}
where we assumed identical statistics for all branches only in the
last line and $P\left(x\right)$ is the marginal distribution of $P\left(x_{1},\dots,x_{B}\right)$.
Similar to App.~\ref{sec:feff_gauss} we defined
\begin{eqnarray}
P_{\mathrm{NL},u\mid x} & := & \frac{1}{2}\mathrm{erfc}\left(\frac{\theta-\mathrm{E}\left[u\mid x\right]}{\sqrt{2\mathrm{Var}\left[u\mid x\right]}}\right),\\
C_{\mathrm{NL},u\mid x} & := & \sqrt{\frac{\mathrm{Var}\left[u\mid x\right]}{2\pi}}\exp\left(-\frac{\left(\theta-\mathrm{E}\left[u\mid x\right]\right)^{2}}{2\mathrm{Var}\left[u\mid x\right]}\right),
\end{eqnarray}
with
\begin{eqnarray}
\mathrm{E}\left[u\mid x\right] & = & x\mathrm{E}\left[w\right],\\
\mathrm{Var}\left[u\mid x\right] & = & x\mathrm{Var}\left[w\right].
\end{eqnarray}
The second moment of $F$ is given by

\begin{eqnarray}
\mathrm{E}\left[F^{2}\right] & = & \int_{-\infty}^{\infty}\mathrm{d}u_{1}\dots\int_{-\infty}^{\infty}\mathrm{d}u_{B}\left(f\left(u_{1}\right)+\dots+f\left(u_{B}\right)\right)^{2}P\left(u_{1},\dots,u_{B}\right)\nonumber \\
 & = & \sum_{x_{1}=1}^{S}\dots\sum_{x_{B}=1}^{S}P\left(x_{1},\dots,x_{B}\right)\sum_{b=1}^{B}\int_{-\infty}^{\infty}\mathrm{d}u_{b}f^{2}\left(u_{b}\right)P\left(u_{b}\mid x_{b}\right)\nonumber \\
 &  & +\sum_{x_{1}=1}^{S}\dots\sum_{x_{B}=1}^{S}P\left(x_{1},\dots,x_{B}\right)\sum_{c=1}^{B}\sum_{b\neq c}^{B}\int_{-\infty}^{\infty}\mathrm{d}u_{b}\int_{-\infty}^{\infty}\mathrm{d}u_{c}f\left(u_{b}\right)f\left(u_{c}\right)P\left(u_{b}\mid x_{b}\right)P\left(u_{c}\mid x_{c}\right)\nonumber \\
 & = & \sum_{x=1}^{S}P\left(x\right)B\left[P_{\mathrm{NL},u\mid x}D^{2}+\left(1-P_{\mathrm{NL},u\mid x}\right)\left(\mathrm{E}^{2}\left[u\mid x\right]+\mathrm{Var}\left[u\mid x\right]\right)-C_{\mathrm{NL},u\mid x}\left(\mathrm{E}\left[u\mid x\right]+\theta\right)\right]\nonumber \\
 &  & +\sum_{x=1}^{S}\sum_{y=1}^{S}P\left(x,y\right)\left(B^{2}-B\right)\left[P_{\mathrm{NL},u\mid x}D+\left(1-P_{\mathrm{NL},u\mid x}\right)\mathrm{E}\left[u\mid x\right]-C_{\mathrm{NL},u\mid x}\right]\nonumber \\
 &  & \,\,\,\,\,\,\,\,\,\,\,\,\,\,\,\,\,\,\,\,\,\,\,\,\,\,\,\,\,\,\,\,\,\,\,\,\,\,\,\,\,\,\,\,\,\,\,\,\,\,\,\,\,\,\,\:\cdot\left[P_{\mathrm{NL},u\mid y}D+\left(1-P_{\mathrm{NL},u\mid y}\right)\mathrm{E}\left[u\mid y\right]-C_{\mathrm{NL},u\mid y}\right],
\end{eqnarray}
where the assumption of identical branch statistics was used in the
last step and $P\left(x,y\right)$is the marginal distribution of
$P\left(x_{1},\dots,x_{B}\right)$.

Analogously, we may compute the exact average number $\mathrm{E}\left[k\right]$
of nonlinear branches. Similar to App.~\ref{sec:feff_gauss}, we
use a step function to count the number $k$ of branches in the nonlinear
regime,
\begin{eqnarray}
\mathrm{E}\left[k\right] & = & \int_{-\infty}^{\infty}\mathrm{d}u_{1}\dots\int_{-\infty}^{\infty}\mathrm{d}u_{B}\left(\mathrm{step}\left(u_{1}-\theta\right)+\dots+\mathrm{step}\left(u_{B}-\theta\right)\right)P\left(u_{1},\dots,u_{B}\right)\nonumber \\
 & = & \sum_{x=1}^{S}P\left(x\right)BP_{\mathrm{NL},u\mid x}.
\end{eqnarray}
The second moment yields
\begin{eqnarray}
\mathrm{E}\left[k^{2}\right] & = & \int_{-\infty}^{\infty}\mathrm{d}u_{1}\dots\int_{-\infty}^{\infty}\mathrm{d}u_{B}\left(\mathrm{step}\left(u_{1}-\theta\right)+\dots+\mathrm{step}\left(u_{B}-\theta\right)\right)^{2}P\left(u_{1},\dots,u_{B}\right)\nonumber \\
 & = & \sum_{x=1}^{S}P\left(x\right)BP_{\mathrm{NL},u\mid x}+\sum_{x=1}^{S}\sum_{y=1}^{S}P\left(x,y\right)\left(B^{2}-B\right)P_{\mathrm{NL},u\mid x}P_{\mathrm{NL},u\mid y}.
\end{eqnarray}
The exact expressions for $\mathrm{E}\left[F\right]$ and $\mathrm{Std}\left[F\right]=\sqrt{\mathrm{E}\left[F^{2}\right]-\mathrm{E}^{2}\left[F\right]}$
as well as $\mathrm{E}\left[k\right]$ and $\mathrm{Std}\left[k\right]=\sqrt{\mathrm{E}\left[k^{2}\right]-\mathrm{E}^{2}\left[k\right]}$
are compared to simulation results and the Gaussian approximation
of $F$ (see App.~\ref{sec:feff_gauss}) in Fig.~\ref{fig:feff}.

\section{Effective somatic input approximation for neurons with multiple layers
of dendritic branches\label{sec:Effective-input-approximation}}

\begin{figure}[!t]
\noindent \begin{centering}
\includegraphics[width=0.75\textwidth]{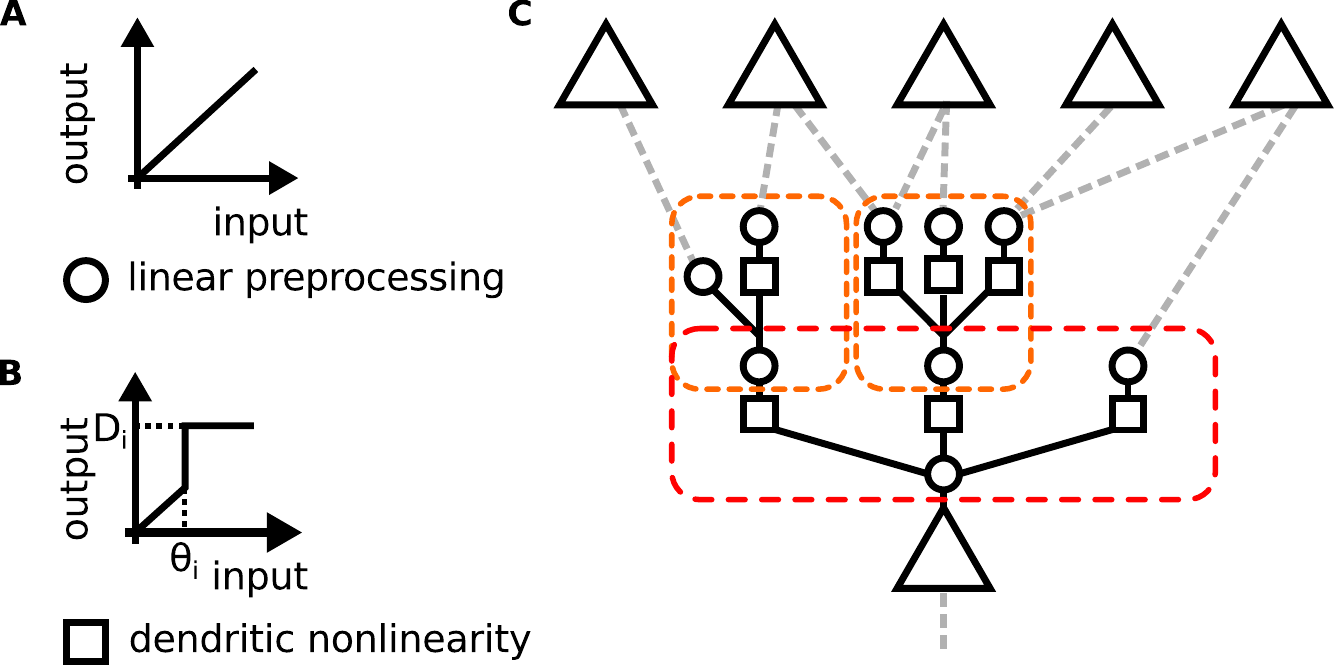}
\par\end{centering}

\protect\caption{\label{fig:multilayer}\textbf{Neuron model with multiple layers of
dendritic branches and non-additive processing.} Similar to Fig.~\ref{fig:model},
different steps of linear \textbf{(A)} and non-additive dendritic
input processing \textbf{(B)} are represented by circles and squares,
respectively. A neuron with two levels of non-additive dendritic branches
is displayed in \textbf{(C)}. As exemplarily shown, the numbers of
sub-branches may vary (at each level and between branches), and there
may be synaptic input to linear and nonlinear dendrites at all levels.}
\end{figure}

The approximation for the somatic input $F$ (Eqs.~\eqref{eq:meanF}-\eqref{eq:varF})
may be readily employed to compute the somatic input for more complex
dendritic arbors and multiple steps of non-additive dendritic processing
(Fig.~\ref{fig:multilayer}). For this, whenever necessary, we assume
that the distribution of inputs may be approximated by the maximum
entropy distribution for given mean and variance, i.e.~by a normal
distribution (cf.~discussion of Eqs.~\eqref{eq:meanU}-\eqref{eq:covU}).
For simplicity of presentation, we assume that the number of sub-branches
$B_{l}$ at a level $l\in\left\{ 1,\dots,L\right\} $ is identical
for all branches $b_{l}\in\left\{ 1,\dots,B_{l}\right\} $ at the
level, that inputs arrive only at the terminal branches and sufficiently
synchronously, and that the distribution of synapses is identical
across the terminal branches. To compute the neuronal input for such
a neuron, we may start from the soma and recursively work towards
the terminal branches: The first moments of the effective somatic
input are given by 

\begin{eqnarray}
\mathrm{E}\left[F\right] & = & B_{1}DP_{\mathrm{NL}}+B_{1}\left(1-P_{\mathrm{NL},1}\right)E\left[u_{1}\right]-B_{1}C_{\mathrm{NL},1},\\
P_{\mathrm{NL},1} & = & P_{\mathrm{NL}}\left(\mathrm{E}\left[u_{1}\right],\mathrm{Var}\left[u_{1}\right]\right),\\
C_{\mathrm{NL},1} & = & C_{\mathrm{NL}}\left(\mathrm{E}\left[u_{1}\right],\mathrm{Var}\left[u_{1}\right]\right),\label{eq:multi_1}
\end{eqnarray}
as defined in Eqs.~\eqref{eq:meanF}-\eqref{eq:cnl} and $\mathrm{Var}\left[F\right]$
can be computed similarly, cf.~Eq.~\eqref{eq:varF}. The appearing
mean input per branch $\mathrm{E}\left[u_{1}\right]$ and its variance
$\mathrm{Var}\left[u_{1}\right]$ are now given by an analogous approximation
that captures the non-additive processing of the preceding layer.
Indeed, the mean input per branch $E\left[u_{l-1}\right]$ for any
layer $l-1$, $L\geq l>1$, is given by
\begin{eqnarray}
\mathrm{E}\left[u_{l-1}\right] & = & B_{l}DP_{\mathrm{NL}}+B_{l}\left(1-P_{\mathrm{NL},l}\right)E\left[u_{l}\right]-B_{l}C_{\mathrm{NL},l},\\
P_{\mathrm{NL},l} & = & P_{\mathrm{NL}}\left(\mathrm{E}\left[u_{l}\right],\mathrm{Var}\left[u_{l}\right]\right),\\
C_{\mathrm{NL},l} & = & C_{\mathrm{NL}}\left(\mathrm{E}\left[u_{l}\right],\mathrm{Var}\left[u_{l}\right]\right),\label{eq:multi_l}
\end{eqnarray}
and $\mathrm{Var}\left[u_{l-1}\right]$ analogously, cf.~Eq.~\eqref{eq:varF}.
In this nomenclature, $\mathrm{E}\left[u_{0}\right]=\mathrm{E}\left[F\right]$
and $\mathrm{Var}\left[u_{0}\right]=\mathrm{Var}\left[F\right]$ capture
the somatic input $F$, and $l=L$ indexes the layer of terminal branches
which receives the synaptic input so that $E\left[u_{L}\right]=E\left[u\right]$
and $Var\left[u_{L}\right]=Var\left[u\right]$ (cf.~Eqs.~\eqref{eq:meanU}-\eqref{eq:covU}).
We note that one can also introduce factors implementing branch coupling
strengths at this point. 

\begin{figure}[!t]
\noindent \begin{centering}
\includegraphics[width=0.48\textwidth]{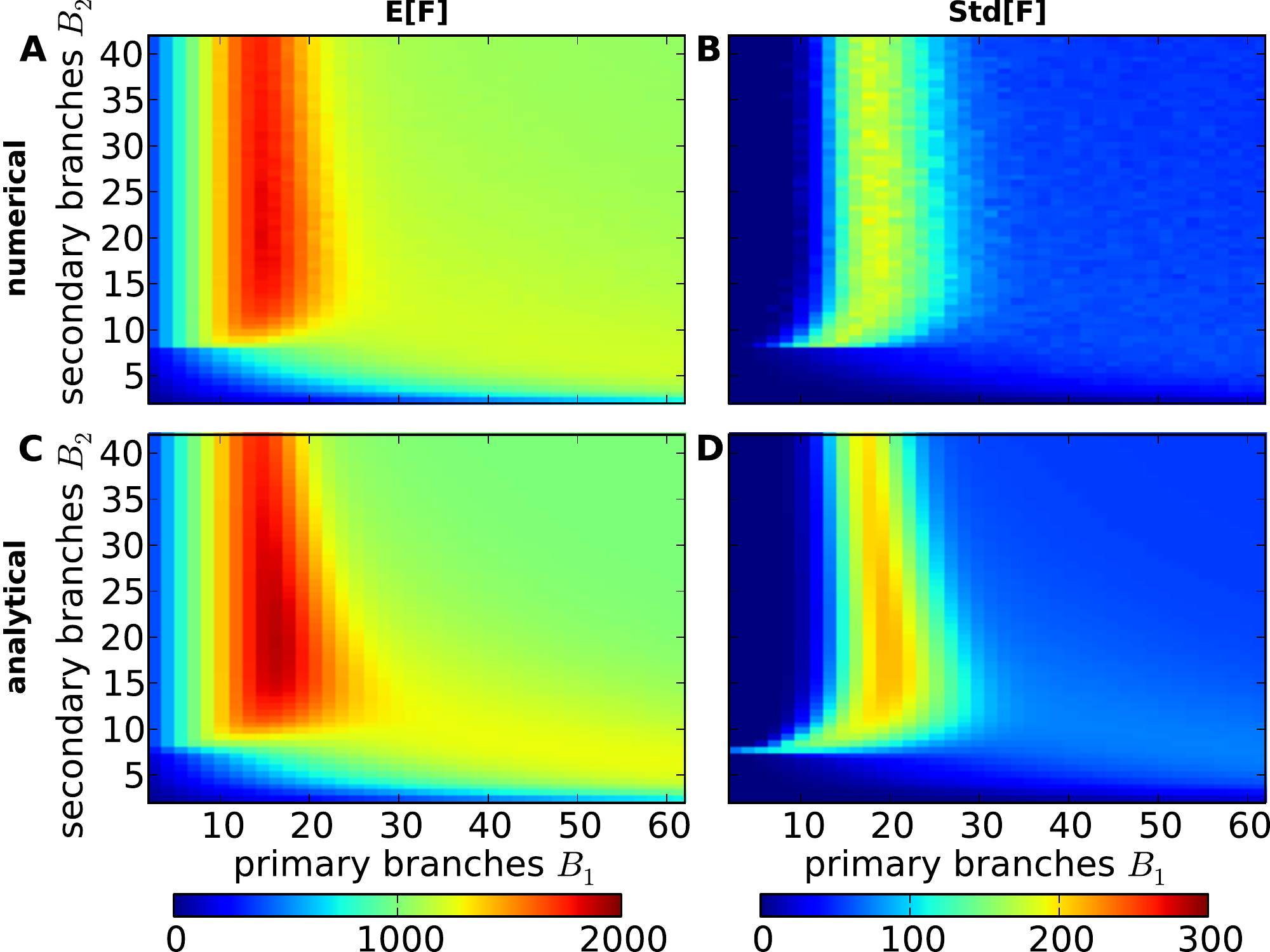}
\par\end{centering}

\protect\caption{\label{fig:twolayer}\textbf{Somatic input for a neuron model with
two layers of non-additive dendritic branches. }Parameters are $\theta_{1}=100$,
$D_{1}=200$, $\theta_{2}=5$, $D_{2}=10$, $\mathrm{E}\left[w\right]=1$,
$\mathrm{Var}\left[w\right]=2$, and $S=1000$, synaptic input arrives
at the terminal branches. Numerical results for the mean somatic input
\textbf{(A)} and its standard deviation\textbf{ (B) }for different
numbers of branches on the first ($B_{1}$) and the second ($B_{2}$)
level are obtained from $400$ realizations of multinomially distributed
synapses. The corresponding analytical results from Eqs.~\eqref{eq:multi_1}
and \eqref{eq:multi_l} are highlighted in \textbf{(C)} and \textbf{(D)}.
Analytical and numerical results agree well. Input is largest for
intermediate numbers of branches $B_{1,\mathrm{opt}}$ and $B_{2,\mathrm{opt}}$.}
\end{figure}

Fig.~\ref{fig:twolayer} shows good agreement of our analytical predictions
with simulation results, both for the mean somatic input $\mathrm{E}\left[F\right]$
and its standard deviation $\mathrm{Std}\left[F\right]$. Dendritic
thresholds and dendritic spike strengths in layer $l=1$ were increased
to compensate for large input strengths in absence of branch coupling
factors. We note that, as for the single-layered neuron, the input
is largest for intermediate numbers of branches $B_{1,\mathrm{opt}}$
and $B_{2,\mathrm{opt}}$ (cf.~discussion of Fig.~\ref{fig:feff}).

The derivation can be directly generalized to cover non-identical
branches, i.e.~branches with different probabilities for the formation
of synapses, different numbers of sub-branches (cf.~App.~\ref{sec:feff_gauss}
and \ref{sec:feff_exact}), linear branches, and additional external
inputs on intermediate level branches. To incorporate linear branches
we may set the dendritic threshold to infinity, to incorporate external
inputs to intermediate level dendrites, we can add an additional,
linear input branch to the considered dendrite, where the external
inputs arrive. We may thus conclude that our approach covers arbitrary
tree-like dendritic structures.

\section{Mean and variance of the input per branch in a network with dendritic
branches\label{sec:distU}}

We now derive the expected input to branch $b$ of neuron $n$ in
the extended Hopfield model when the states of the neurons are fixed
and the average is taken over the weight distribution. By construction,
we have
\begin{eqnarray}
\mathrm{E}\left[u_{n,b}\right] & = & \mathrm{E}\left[\sum_{m=1}^{N}w_{n,b,m}v_{m}\right]\nonumber \\
 & = & \sum_{m=1}^{N}\mathrm{E}\left[w_{n,b,m}\right]v_{m}\nonumber \\
 & = & \sum_{m=1}^{N}B^{-1}w_{n,m}v_{m}\nonumber \\
 & = & B^{-1}u_{n},
\end{eqnarray}
so that the mean input to the branch depends only on the field $u_{n}$
of the classical Hopfield model. The choice $\mathrm{E}\left[w_{n,b,m}\right]=B^{-1}w_{n,m}$
in line three with $w_{n,m}$ given by Eq.~\eqref{eq:hebb} is justified
by assuming Hebbian learning. The variance of the input per branch
is given by
\begin{eqnarray}
\mathrm{Var}\left[u_{n,b}\right] & = & \mathrm{Var}\left[\sum_{m=1}^{N}w_{n,b,m}v_{m}\right]\nonumber \\
 & = & \sum_{m=1}^{N}v_{m}^{2}\mathrm{Var}\left[w_{n,b,m}\right]+\sum_{m\neq k}^{N}v_{m}v_{k}\mathrm{Cov}\left[w_{n,b,m},w_{n,b,k}\right]\nonumber \\
 & = & \sum_{m=1}^{N}\mathrm{Var}\left[w_{n,b,m}\right]+0\nonumber \\
 & = & \sum_{m=1}^{N}w_{n,m}^{2}B^{-2}\mathrm{Var}\left[w\right]\nonumber \\
 & = & \mathrm{Var}\left[w\right]B^{-2}N\mathrm{E}\left[w_{n,m}^{2}\right]\nonumber \\
 & = & \mathrm{Var}\left[w\right]B^{-2}N\left(\mathrm{Var}\left[w_{n,m}\right]+\mathrm{E}^{2}\left[w_{n,m}\right]\right),
\end{eqnarray}
where we used the independence of the $w_{n,b,m}$ in the third line.
In the fourth line, we employed $\mathrm{Var}\left[w_{n,b,m}\right]=w_{n,m}^{2}B^{-2}\mathrm{Var}\left[w\right]$
with a parameter $\mathrm{Var}\left[w\right]$, cf.~the discussion
preceding Eqs.~\eqref{eq:meanUn}-\eqref{eq:varUn}. For the Hopfield
network with $P$ random patterns to be stored (Eq.~\eqref{eq:hebb}),
$\mathrm{E}\left[w_{n,m}\right]=0$ and $\mathrm{Var}\left[w_{n,m}\right]=PN^{-2}=N^{-1}\alpha$
(with the load $\alpha=PN^{-1}$) since $w_{n,m}$ is a sum of $P$
contributions $\pm N^{-1}$ with equal probabilities. Finally, in
our extended Hopfield model, the correlation of input between different
branches vanishes,
\begin{eqnarray}
\mathrm{Cov}\left[u_{n,b},u_{n,c}\right] & = & \mathrm{Cov}\left[\sum_{m=1}^{N}u_{n,b,m}v_{m},\sum_{k=1}^{N}u_{n,c,k}v_{k}\right]\nonumber \\
 & = & \sum_{m=1}^{N}\sum_{k=1}^{N}v_{m}v_{k}\mathrm{Cov}\left[u_{n,b,m},u_{n,c,k}\right]\nonumber \\
 & = & 0,
\end{eqnarray}
where $b\neq c$. In the third line, we used that for $m=k$ the $u_{n,b,m}$
are independently chosen from (the same) Gaussian distributions with
means $B^{-1}w_{n,m}$ and variances $B^{-2}w_{n,m}^{2}\mathrm{Var}\left[w\right]$
and for $m\neq k$ they are independently chosen from their respective
(in general different) distributions. Because the input across branches
is uncorrelated, we may employ the results for the somatic input we
derived for binomially distributed active synapses (Eqs.~\eqref{eq:meanF}
and \eqref{eq:varFbino}).

\section{Convergence of a Hopfield network with dendritic nonlinearities\label{sec:conv}}

Here, we show that the dynamics of the deterministic Hopfield network
with nonlinear dendrites are equivalent to those of the classical
Hopfield model with reduced threshold. In particular, network convergence
is guaranteed. Assuming that $\bar{F}$ (Eq.~\eqref{eq:F_explicitly_Hopfield})
is strictly monotonically increasing and therefore also invertible,
we find that
\begin{eqnarray}
v_{n}\left(t+1\right)=\mathrm{sign}\left(\bar{F}\left(u_{n}\left(t\right)\right)-\Theta\right) & = & \mathrm{sign}\left(u_{n}\left(t\right)-\vartheta\right),
\end{eqnarray}
with the effective threshold
\begin{eqnarray}
\vartheta & = & \bar{F}^{-1}\left(\Theta\right),\label{eq:app_conv_uc}
\end{eqnarray}
see also Fig.~\ref{fig:thresred}. This update rule is equivalent
to the conventional one with threshold $\vartheta$. Hence, the energy
function $E_{\mathrm{NL}}$ of the system is obtained from $E_{\mathrm{L}}$
(Eq.~\eqref{eq:Elin}) by substituting $\Theta$ by $\vartheta$,
\begin{eqnarray}
E_{\mathrm{NL}}\left(v_{1}\left(t\right),\dots,v_{N}\left(t\right)\right) & = & -\frac{1}{2}\sum_{n=1}^{N}\sum_{m=1}^{N}w_{n,m}v_{n}\left(t\right)v_{m}\left(t\right)+\sum_{m=1}^{N}\vartheta v_{m}\left(t\right).
\end{eqnarray}
For symmetric couplings, $w_{n,m}=w_{m,n}$, $E_{\mathrm{NL}}$ is
monotonically decreasing and the system converges to a steady state.
More explicitly, by assuming that neuron $n$ is updated we have
\begin{eqnarray}
E_{\mathrm{NL}}\left(t+1\right)-E_{\mathrm{NL}}\left(t\right) & = & -\left(v_{n}\left(t+1\right)-v_{n}\left(t\right)\right)\left(u_{n}\left(t\right)-\vartheta\right)\leq0.
\end{eqnarray}
Equality holds only for $v_{n}\left(t+1\right)=v_{n}\left(t\right)$
or $u_{n}\left(t\right)=\vartheta$ where the latter implies $v_{n}\left(t+1\right)=1$.
Therefore, the energy $E_{\mathrm{NL}}$ either decreases in time
or remains constant only if the state of the network does not change
or the updated neuron $n$ is set to $v_{n}=1$. Since the energy
is bounded from below due to
\begin{eqnarray}
\left|E_{\mathrm{NL}}\right| & \leq & \frac{1}{2}\sum_{n=1}^{N}\sum_{m=1}^{N}\left|w_{n,m}\right|+N\vartheta,
\end{eqnarray}
the network converges to a stable state which is given by a minimum
in the energy landscape $E_{\mathrm{NL}}\left(v_{1},\dots,v_{N}\right)$.
Thus, the effective nonlinearity reduces the neuronal threshold to
$\vartheta\leq\Theta$ as compared to linear input summation (Eq.~\eqref{eq:Elin})
but maintains network convergence.

\begin{figure}[!t]
\noindent \begin{centering}
\includegraphics[width=0.48\textwidth]{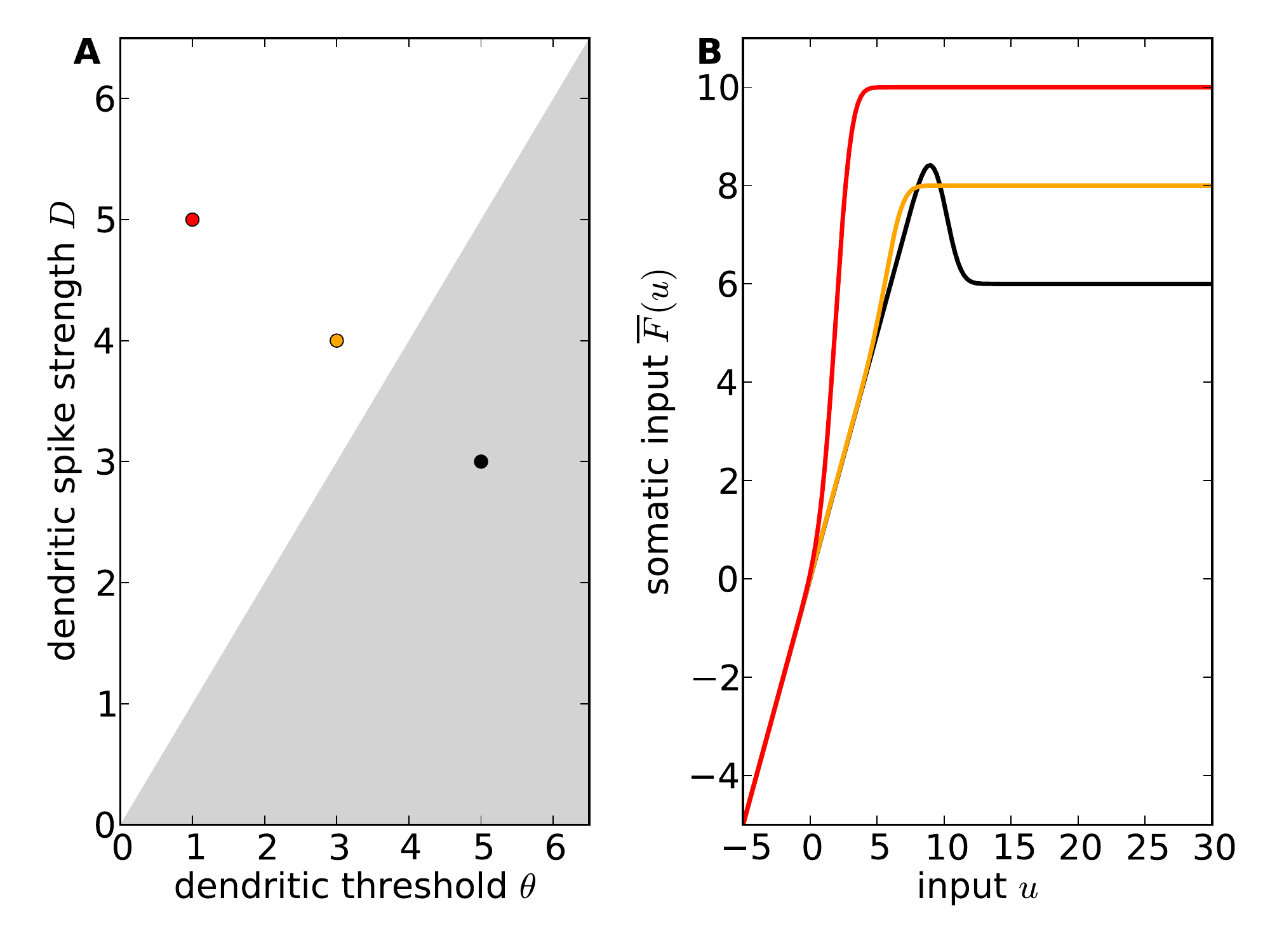}
\par\end{centering}

\protect\caption{\label{fig:mono}\textbf{Monotonicity of the effective somatic input
$\bar{F}$.} Panel (\textbf{A}) indicates the monotonicity of $\bar{F}$
(Eq.~\eqref{eq:F_explicitly_Hopfield}) for $B=2$, $PN^{-1}\mathrm{Var}\left[w\right]=0.8$
(cf.~caption of Fig.~\ref{fig:thresred}) with varying dendritic
thresholds $\theta$ and dendritic spike strengths $D$. It displays
two regions, one where $\bar{F}$ is strictly monotonic (white) and
one where it is non-monotonic (gray). They are separated by $D=\theta$.
Panel (\textbf{B}) shows the monotonic or non-monotonic shape of $\bar{F}$
for different dendritic parameters $\left(\theta,D\right)\in\left\{ \left(5,3\right),\left(3,4\right),\left(1,5\right)\right\} $
(black, orange, red).}
\end{figure}

For $\vartheta$ to be uniquely defined, the dendritic nonlinearities
have to be strong enough, $BD>\Theta$, so that $\bar{F}$ intersects
the constant function $\Theta$ (Fig.~\ref{fig:thresred}). Analytical
calculations show that for $D>\theta$ the transfer function $\bar{F}$
is strictly monotonic (cf.~Fig.~\ref{fig:mono}). Since experiments
demonstrate supralinear dendritic amplification, e.g., with thresholds
of $\theta\approx3.8\,\mathrm{mV}$ and spike amplitudes of $D\approx10\,\mathrm{mV}$
\cite{Ariav2003}, this parameter regime is biologically plausible.

\section{Asymmetric couplings and convergence of a Hopfield network with nonlinear
dendrites\label{sec:asym}}

The couplings of the classical Hopfield network are symmetric, $w_{n,m}=w_{m,n}$,
so that convergence is guaranteed by a Lyapunov function (Eq.~\eqref{eq:Elin}).
In the extended model, we argued that the coupling weights $w_{n,b,m}$
to the dendritic branches obey $\mathrm{E}\left[w_{n,b,m}\right]=B^{-1}w_{n,m}$
or, equivalently, $B\mathrm{E}\left[w_{n,b,m}\right]=w_{n,m}$ due
to Hebbian learning (Eq.~\eqref{eq:meanUn}). To account for fluctuations
in the learned weights we assumed a variance $\mathrm{Var}\left[w_{n,b,m}\right]=w_{n,m}^{2}B^{-2}\mathrm{Var}\left[w\right]$
(Eq.~\eqref{eq:varUn}). In a particular network realization with
a finite number of branches and fluctuations, the dendritic weights
do therefore not sum up to the expected Hebbian weight precisely,
\begin{eqnarray}
\sum_{b=1}^{B}w_{n,b,m} & =: & w_{n,m}^{'}\neq w_{n,m}.
\end{eqnarray}
Generally, we have $w_{n,m}^{'}\neq w_{m,n}^{'}$ and the magnitude
of the deviation from symmetric couplings is determined by $\mathrm{Var}\left[w\right]$.
To check if our analytical calculations are applicable despite larger
asymmetries, we redo the simulations from the main part of the paper
for larger $\mathrm{Var}\left[w\right]$.

\begin{figure}[!t]
\noindent \begin{centering}
\includegraphics[width=0.48\textwidth]{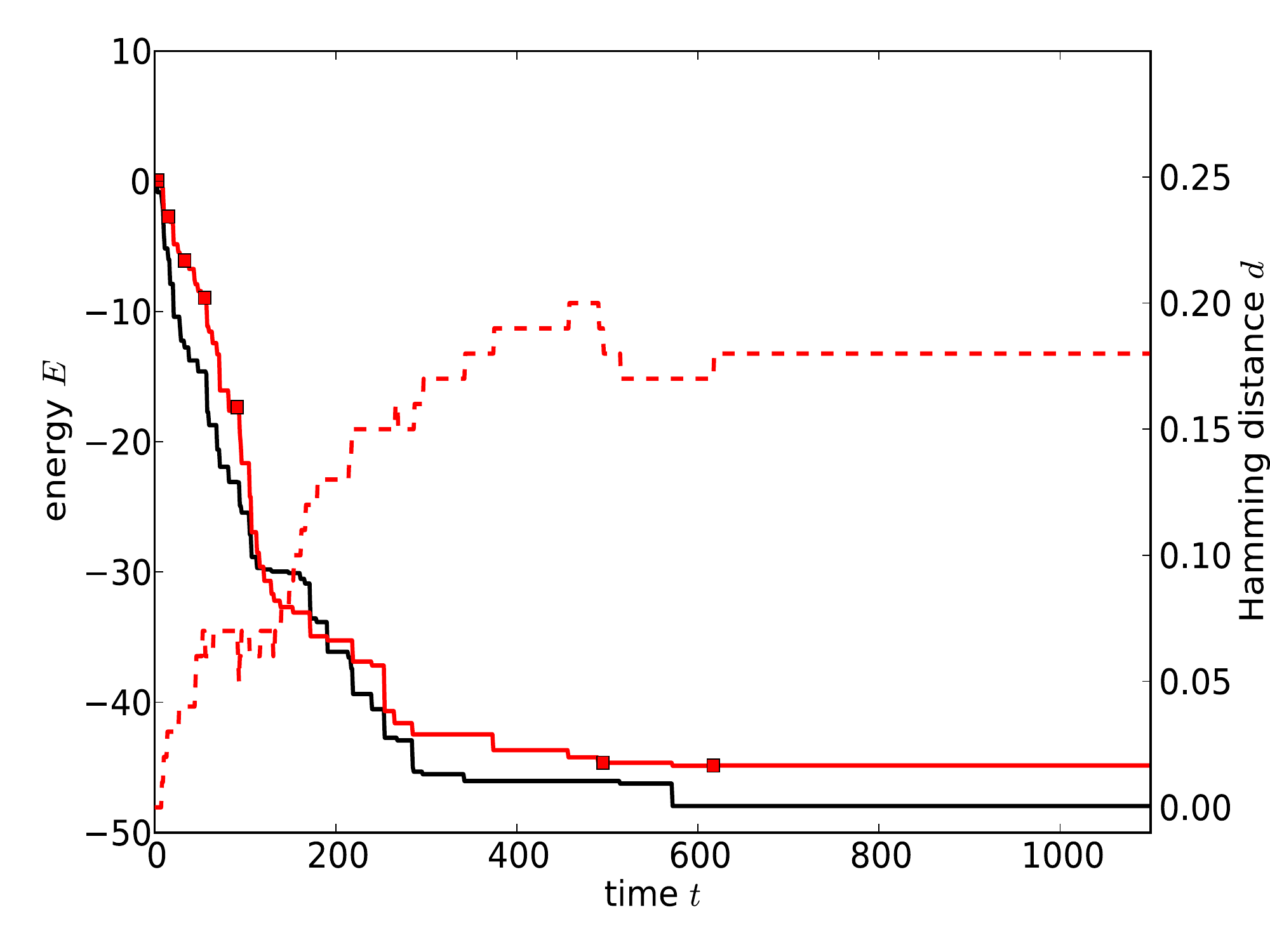}
\par\end{centering}

\protect\caption{\label{fig:asym}\textbf{Network convergence despite more asymmetric
couplings.} Parameters are the same as in Fig.~\ref{fig:conv}, with
$\mathrm{Var}\left[w\right]=0.5$. Simulations are performed with
identical initial states, topology and order of updates. As the linear
Hopfield model is not affected by $\mathrm{Var}\left[w\right]$, its
energy (solid black) decreases monotonically and reaches a fixed point.
For the extended Hopfield model with non-additive dendrites and more
asymmetric couplings, the energy decreases (solid red) with rare events
of increasing energy (red squares) due to deviations from the mean-field
approach or asymmetric couplings (cf.~discussion of Eq.~\eqref{eq:Enonlin}).
The convergence of the system is preserved (checked for $1000$ runs).
The Hamming distance $d=\frac{1}{2N}\sum_{n=1}^{N}\left|v_{n}-v_{n}^{'}\right|$
(dashed red) between the systems shows that they settle into different
attractors.}
\end{figure}

\begin{figure}[!t]
\noindent \begin{centering}
\includegraphics[width=0.48\textwidth]{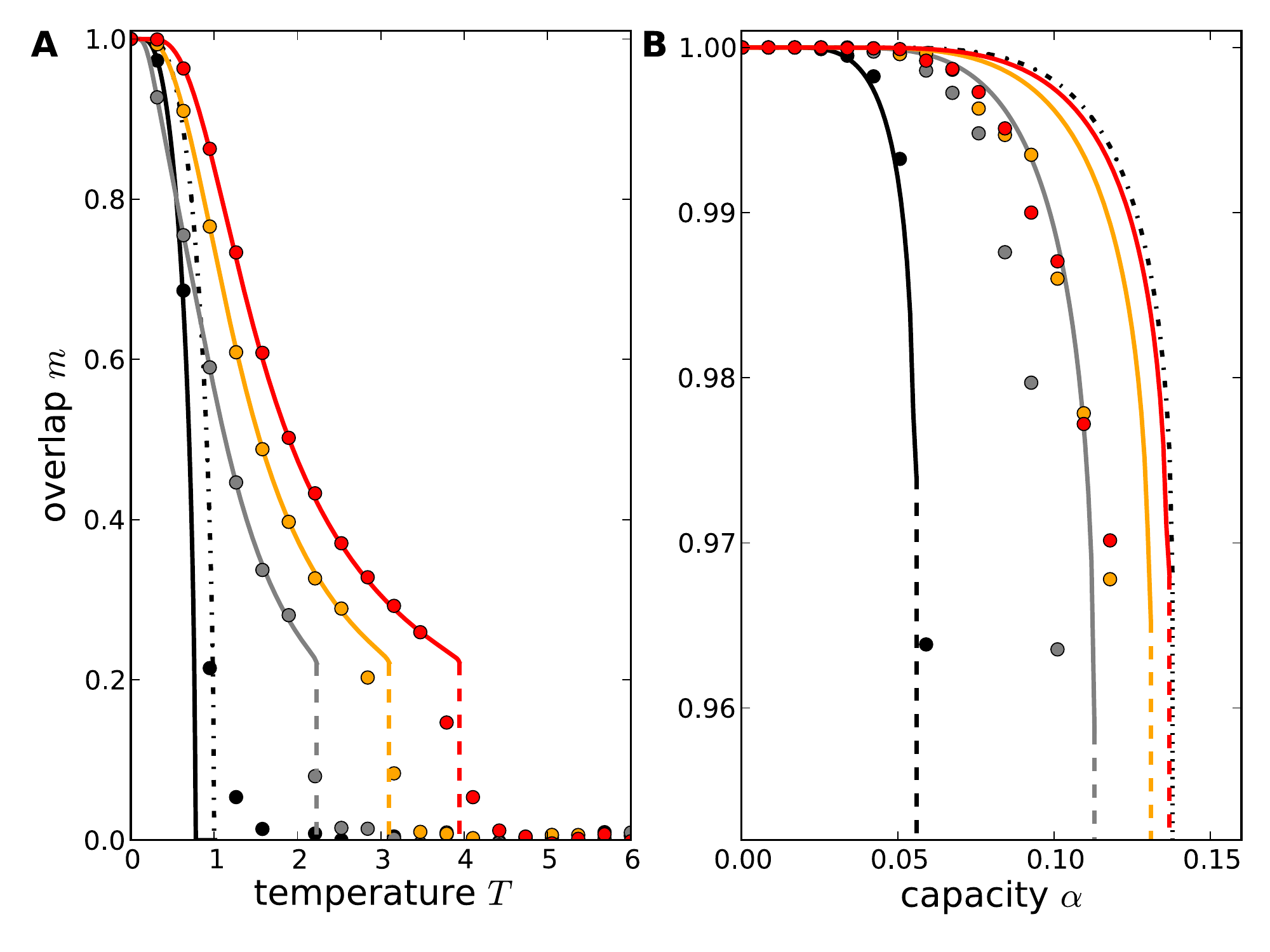}
\par\end{centering}

\protect\caption{\label{fig:asmag}\textbf{Memory performance of the stochastic Hopfield
model with nonlinear dendritic branches and more asymmetric couplings.}
Parameters are the same as in Fig.~\ref{fig:mag}, with $\mathrm{Var}\left[w\right]=0.5$.
The figure compares analytical results (solid lines) for linear summation
(black) and nonlinear summation with dendritic spike strengths $D\in\left\{ 0.4,0.6,0.8\right\} $
(gray, orange, red) and symmetric couplings to simulation results
(circles) for asymmetric couplings. Results for linear input summation
with $\Theta=0$ are included for comparison (dash-dotted black).
Panel (\textbf{A}) shows the overlap $m$ versus the temperature $T$
for a small load $\alpha=N^{-1}\approx0$. The more asymmetric couplings
change $m$ only slightly and the simulation results agree well with
the analytics of the symmetric couplings. Panel (\textbf{B}) shows
$m$ versus $\alpha$ for $T=0$. The stronger asymmetries impair
the memory function and the overlap $m$ displays a drop at lower
loads $\alpha$ compared to the symmetric case.}
\end{figure}

Our simulations indicate that although convergence of the extended
deterministic Hopfield network is not guaranteed by a Lyapunov function
for asymmetric couplings, it reaches a fixed point as shown exemplarily
by Fig.~\ref{fig:asym} and confirmed for $1000$ runs (not shown)
also for larger asymmetries $\mathrm{Var}\left[w\right]$. To study
the impact of asymmetric couplings on the memory performance of the
extended stochastic Hopfield network, we repeat the simulations shown
in Fig.~\ref{fig:mag} for larger $\mathrm{Var}\left[w\right]$.
For a small load $\alpha\approx0$ and non-zero temperatures $T>0$,
the analytical calculations for symmetric couplings agree well with
the simulation results (Fig.~\ref{fig:asmag}A). Yet, in the zero
temperature limit $T=0$, the asymmetries decrease the storage capacity
of the network compared to the symmetric case (Fig.~\ref{fig:asmag}B).

\section{Mean-field calculations for a stochastic Hopfield model with nonlinear
dendritic branches\label{sec:meanfield}}

We now derive the mean-field equations for the overlap $m:=m^{1}=N^{-1}\sum_{n=1}^{N}\xi_{n}^{1}\left\langle v_{n}\right\rangle $
of the network state $\left(v_{1},\dots,v_{N}\right)$ with pattern
$p=1$, i.e.~$\left(\xi_{1}^{1},\dots,\xi_{N}^{1}\right)$. The following
calculations go along those provided in \cite{Geszti1990,hertz1991introduction}.
From
\begin{eqnarray}
\left\langle u_{n}\right\rangle  & = & \sum_{k=1}^{N}w_{n,k}\left\langle v_{k}\right\rangle =\sum_{k=1}^{N}N^{-1}\sum_{p=1}^{P}\xi_{n}^{p}\xi_{k}^{p}\left\langle v_{k}\right\rangle =\sum_{p=1}^{P}\xi_{n}^{p}m^{p}=:u_{n}\label{eq:app_Fu}
\end{eqnarray}
and Eq.~\eqref{eq:meanfield} we find
\begin{eqnarray}
m^{q} & = & N^{-1}\sum_{n=1}^{N}\xi_{n}^{q}\tanh\left[\beta\left(\bar{F}\left(u_{n}\right)-\Theta\right)\right]\nonumber \\
 & = & N^{-1}\sum_{n=1}^{N}\xi_{n}^{q}\xi_{n}^{1}\tanh\left[\beta\xi_{n}^{1}\left(\bar{F}\left(u_{n}\right)-\Theta\right)\right],\label{eq:app_Fmq}
\end{eqnarray}
where we employed the point symmetry of $\tanh\left(x\right)=-\tanh\left(-x\right)$
in the second line. We now assume that the number $P$ of patterns
is large, of order $\mathcal{O}\left(N\right)$. We define the mean
square overlap $r:=\alpha^{-1}\sum_{q\neq1}^{P}\left(m^{q}\right)^{2}$
and assume that the $m^{q}$ are independent, zero-centered random
variables with variance $\alpha rP^{-1}$. Then, the sum $\xi_{n}^{1}\sum_{p\neq1}^{P}\xi_{n}^{p}m^{p}$
can be seen as a Gaussian noise term of variance $\alpha r$ and the
sum $N^{-1}\sum_{n=1}^{N}$ may be treated as an average over this
noise. Since $\xi_{n}^{1}=\pm1$ with equal probabilities, the overlap
with the first pattern is (Eq.~\eqref{eq:app_Fmq})
\begin{eqnarray}
m & = & \frac{1}{2}\int_{-\infty}^{\infty}\frac{dz}{\sqrt{2\pi}}e^{-\frac{1}{2}z^{2}}\tanh\left[\beta\left(\bar{F}\left(m+\sqrt{\alpha r}z\right)-\Theta\right)\right]\nonumber \\
 &  & +\frac{1}{2}\int_{-\infty}^{\infty}\frac{dz}{\sqrt{2\pi}}e^{-\frac{1}{2}z^{2}}\tanh\left[-\beta\left(\bar{F}\left(-m-\sqrt{\alpha r}z\right)-\Theta\right)\right],\label{eq:app_Fm}
\end{eqnarray}
with the effective somatic input $\bar{F}$ given by Eqs.~\eqref{eq:F_explicitly_Hopfield}-\eqref{eq:cnl_explicitly_Hopfield}.
Next,the correlations quantified by $r$ must be determined self-consistently.
We define $\bar{u}_{n}^{q}:=\sum_{p\neq q}^{P}\xi_{n}^{p}m^{p}$ and
because $\xi_{n}^{q}m^{q}=u_{n}-\bar{u}_{n}^{q}$ is small, of order
$\mathcal{O}\left(N^{-1/2}\right)$, we expand Eq.~\eqref{eq:app_Fmq}
into a Taylor series to first order,
\begin{eqnarray}
m^{q} & \approx & N^{-1}\sum_{n=1}^{N}\xi_{n}^{q}\xi_{n}^{1}\tanh\left[\beta\xi_{n}^{1}\left(\bar{F}\left(\bar{u}_{n}^{q}\right)-\Theta\right)\right]\nonumber \\
 &  & +\beta N^{-1}\sum_{n=1}^{N}\left(1-\tanh^{2}\left[\beta\xi_{n}^{1}\left(\bar{F}\left(\bar{u}_{n}^{q}\right)-\Theta\right)\right]\right)\bar{F}^{'}\left(\bar{u}_{n}^{q}\right)m^{q},\label{eq:app_Ftaylor}
\end{eqnarray}
where $f^{'}\left(x_{0}\right)=\left.\frac{df\left(x\right)}{dx}\right|_{x=x_{0}}$
denotes the first derivative of a function $f\left(x\right)$ evaluated
at $x=x_{0}$. Similar to the derivation of Eq.~\eqref{eq:app_Fm},
we approximate $\xi_{n}^{1}\sum_{p\neq1,q}^{P}\xi_{n}^{p}m^{p}$ by
a zero-centered Gaussian distribution with variance $\alpha r$ and
in the second term of Eq.~\eqref{eq:app_Ftaylor} treat the sum $N^{-1}\sum_{n=1}^{N}$
as an average so that
\begin{eqnarray}
m^{q} & = & N^{-1}\sum_{n=1}^{N}\xi_{n}^{q}\xi_{n}^{1}\tanh\left[\beta\xi_{n}^{1}\left(\bar{F}\left(\bar{u}_{n}^{q}\right)-\Theta\right)\right]+\beta Cm^{q},\label{eq:app_Fmqts}
\end{eqnarray}
with
\begin{eqnarray}
C & := & \frac{1}{2}\int_{-\infty}^{\infty}\frac{dz}{\sqrt{2\pi}}e^{-\frac{1}{2}z^{2}}\left(1-\tanh^{2}\left[\beta\left(\bar{F}\left(m+\sqrt{\alpha r}z\right)-\Theta\right)\right]\right)\bar{F}^{'}\left(m+\sqrt{\alpha r}z\right)\nonumber \\
 &  & +\frac{1}{2}\int_{-\infty}^{\infty}\frac{dz}{\sqrt{2\pi}}e^{-\frac{1}{2}z^{2}}\left(1-\tanh^{2}\left[-\beta\left(\bar{F}\left(-m-\sqrt{\alpha r}z\right)-\Theta\right)\right]\right)\bar{F}^{'}\left(-m-\sqrt{\alpha r}z\right),\,\,\,\,\,\,\,\,\,\,\,\,\,\,\label{eq:app_Fc}
\end{eqnarray}
where we took into account again that $\xi_{n}^{1}=\pm1$ with equal
probabilities. Solving Eq.~\eqref{eq:app_Fmqts} for $m^{q}$, squaring
it and averaging over all patterns $q$ yields
\begin{eqnarray}
r & = & \left(1-\beta C\right)^{-2}NP^{-1}\sum_{q\neq1}^{P}N^{-2}\sum_{n,k=1}^{N}\xi_{n}^{q}\xi_{n}^{1}\xi_{k}^{q}\xi_{k}^{1}\nonumber \\
 &  & \cdot\tanh\left[\beta\xi_{n}^{1}\left(\bar{F}\left(\bar{u}_{n}^{q}\right)-\Theta\right)\right]\tanh\left[\beta\xi_{k}^{1}\left(\bar{F}\left(\bar{u}_{k}^{q}\right)-\Theta\right)\right]\nonumber \\
 & = & \left(1-\beta C\right)^{-2}N^{-1}\sum_{n=1}^{N}\tanh^{2}\left[\beta\xi_{n}^{1}\left(\bar{F}\left(\bar{u}_{n}^{2}\right)-\Theta\right)\right].
\end{eqnarray}
Here, we used that the arguments of the $\tanh$ are independent of
$q$ and in the average $P^{-1}\sum_{q\neq1}^{P}$ only terms $n=k$
survive. We set $\bar{u}_{n}^{q}=\bar{u}_{n}^{2}$ without loss of
generality. Employing a Gaussian approximation of the sum $\xi_{n}^{1}\sum_{p\neq1,2}^{P}\xi_{n}^{p}m^{p}$
like in Eq.~\eqref{eq:app_Fmqts} yields
\begin{eqnarray}
r & = & \left(1-\beta C\right)^{-2}s,\label{eq:app_Fr}
\end{eqnarray}
with
\begin{eqnarray}
s & := & \frac{1}{2}\int_{-\infty}^{\infty}\frac{dz}{\sqrt{2\pi}}e^{-\frac{1}{2}z^{2}}\tanh^{2}\left[\beta\left(\bar{F}\left(m+\sqrt{\alpha r}z\right)-\Theta\right)\right]\nonumber \\
 &  & +\frac{1}{2}\int_{-\infty}^{\infty}\frac{dz}{\sqrt{2\pi}}e^{-\frac{1}{2}z^{2}}\tanh^{2}\left[-\beta\left(\bar{F}\left(-m-\sqrt{\alpha r}z\right)-\Theta\right)\right].\label{eq:app_Fs}
\end{eqnarray}
The above Gaussian approximations hold for $\alpha$ of order $\mathcal{O}\left(1\right)$
and smaller. Eqs.~\eqref{eq:app_Fm}, \eqref{eq:app_Fc}, \eqref{eq:app_Fr},
and\eqref{eq:app_Fs} constitute a set of nonlinear, coupled integral
equations for the order parameters $m$, $r$ and $s$ and can be
solved numerically or in limiting cases.

\section{Memory capacity of a stochastic Hopfield network with nonlinear dendritic
branches in the thermodynamic limit for a finite number of patterns\label{sec:lima}}

We now consider the quality of pattern retrieval estimated by the
overlap $m$ in the thermodynamic limit of large $N$ with finitely
many patterns $P$, i.e.~$\alpha\approx0$. Because $m^{q}$, $q\neq1$,
is of order $\mathcal{O}\left(N^{-1/2}\right)$ and $P$ is finite
we may write $u_{n}=\sum_{p=1}^{P}\xi_{n}^{p}m^{p}\approx\xi_{n}^{1}m$.
Starting from Eq.~\eqref{eq:app_Fmq} and using the definition of
the effective somatic input $\bar{F}$ (Eqs.~\eqref{eq:F_explicitly_Hopfield}-\eqref{eq:cnl_explicitly_Hopfield}),
\begin{eqnarray}
m & = & N^{-1}\sum_{n=1}^{N}\tanh\left[\beta\left(1-P_{\mathrm{NL}}\left(u_{n}\right)\right)\xi_{n}^{1}u_{n}+\beta\xi_{n}^{1}\left(BDP_{\mathrm{NL}}\left(u_{n}\right)-BC_{\mathrm{NL}}\left(u_{n}\right)-\Theta\right)\right]\nonumber \\
 & \approx & N^{-1}\sum_{n=1}^{N}\tanh\left[\beta\left(1-P_{\mathrm{NL}}\left(\xi_{n}^{1}m\right)\right)m+\beta\xi_{n}^{1}\left(BDP_{\mathrm{NL}}\left(\xi_{n}^{1}m\right)-BC_{\mathrm{NL}}\left(\xi_{n}^{1}m\right)-\Theta\right)\right]\nonumber \\
 & = & \frac{1}{2}\tanh\left[\beta\left(1-P_{\mathrm{NL}}\left(m\right)\right)m+\beta\left(BDP_{\mathrm{NL}}\left(m\right)-BC_{\mathrm{NL}}\left(m\right)-\Theta\right)\right]\nonumber \\
 &  & +\frac{1}{2}\tanh\left[\beta\left(1-P_{\mathrm{NL}}\left(-m\right)\right)m-\beta\left(BDP_{\mathrm{NL}}\left(-m\right)-BC_{\mathrm{NL}}\left(-m\right)-\Theta\right)\right],\label{eq:app_lima}
\end{eqnarray}
where we use that $\xi_{n}^{1}=\pm1$ with equal probabilities in
the third line. This transcendental equation for $m$ is solved numerically
(Fig.~\ref{fig:mag}).

\begin{figure}[!t]
\noindent \begin{centering}
\includegraphics[width=0.48\textwidth]{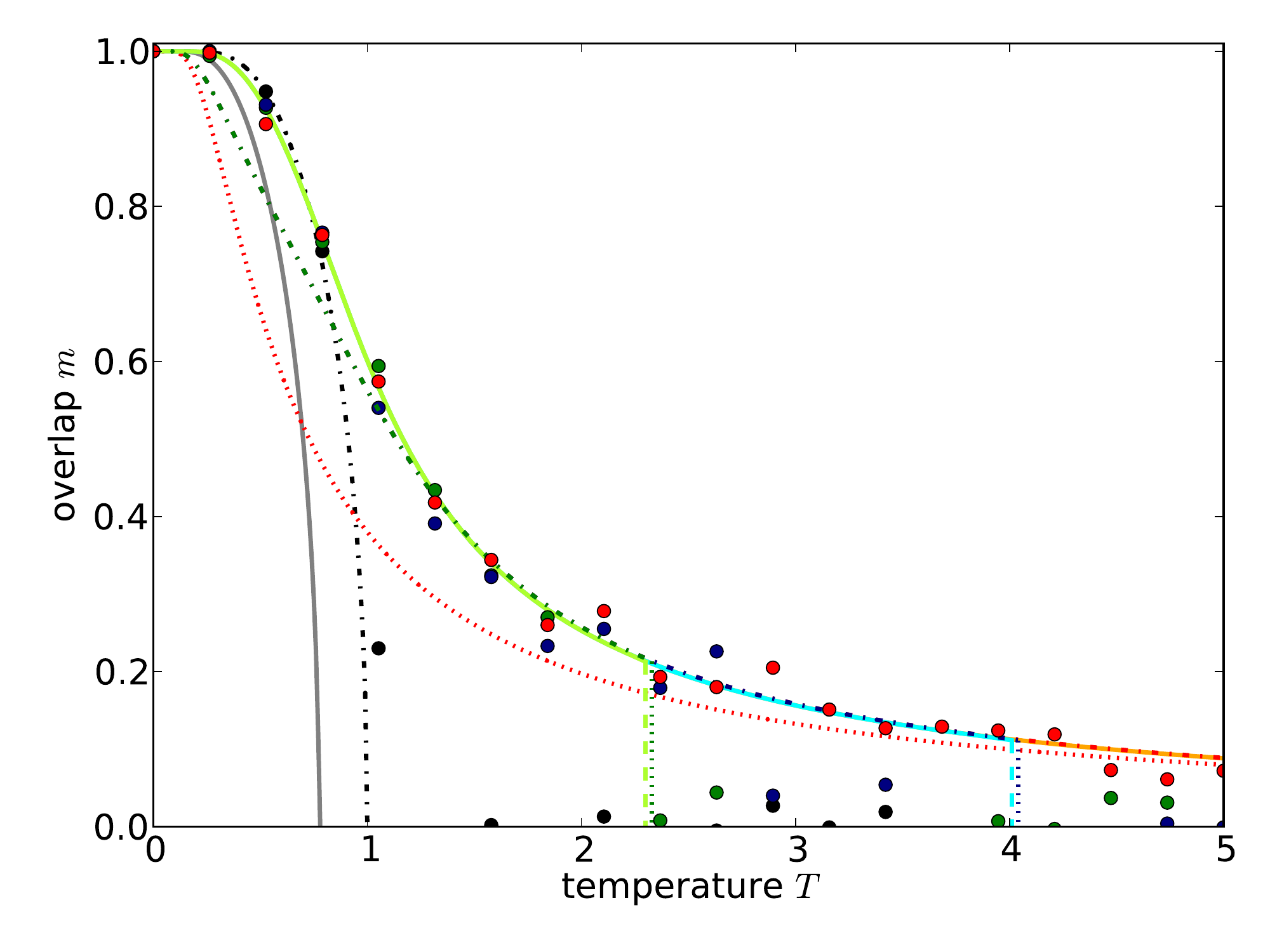}
\par\end{centering}

\protect\caption{\label{fig:app_thc}\textbf{Critical behavior of the network for few
patterns $\alpha\approx0$ and varying neuronal thresholds $\Theta$
and dendritic thresholds $\theta$.} Parameters are the same as in
Fig.~\ref{fig:mag}, with $D=0.4$ and different dendritic thresholds
$\theta\in\left\{ \infty,0.1,0.05,0\right\} $. Simulation results
(circles) and analytical results (dash-dotted lines) for vanishing
neuronal thresholds $\Theta=0$ are compared to the analytical results
for $\Theta=0.4$ (solid lines; cf.~Fig.~\ref{fig:mag}). $\theta=\infty$
(dash-dotted black and solid gray) corresponds to the linear Hopfield
model. Decreasing the dendritic threshold $\theta$ increases the
critical temperature ($\theta=0.1$ in green and lime, $\theta=0.05$
in blue and cyan). For $\theta=0$ (red and orange), there is no phase
transition anymore as $m\left(T\right)$ asymptotically follows Eq.~\eqref{eq:app_thc}
(dotted red).}
\end{figure}

It is shown in the main text that dendritic nonlinearities elevate
the critical temperature $T_{c}$ above which retrieval fails (Fig.~\ref{fig:mag}).
The increased critical temperature $T_{c}$ may result partially from
the effectively reduced neuronal threshold $\vartheta\leq\Theta$
(Eq.~\eqref{eq:uc}). To exclude this effect, we study the impact
of the nonlinearity on the critical temperature $T_{c}$ for vanishing
neuronal threshold $\Theta=0$. We find that when changing $\Theta$
from $\Theta=0.4$ to $\Theta=0$ the critical temperature $T_{c}$
is altered only slightly (Fig.~\ref{fig:app_thc}, see differences
between dashed and dotted vertical lines) and the behavior of the
system remains the same.

In contrast, the threshold $\theta$ of the dendritic nonlinearity
has a strong influence on the critical behavior of the network. For
$\theta=\infty$, we reobtain linear input summation (Fig.~\ref{fig:app_thc},
dash-dotted black and solid gray). For finite $0<\theta<\infty$,
the critical temperature $T_{c}$ is increased (Fig.~\ref{fig:app_thc},
green and lime, blue and cyan). For $\theta=0$, there is no phase
transition at all (Fig.~\ref{fig:app_thc}, red and orange) and memory
retrieval at arbitrarily high temperatures $T$ (although with smaller
and smaller overlaps$m$) is possible. This may be understood by assuming
$\theta=0$ and $T=\beta^{-1}\rightarrow\infty$ in Eq.~\eqref{eq:app_lima}.
Then,
\begin{eqnarray}
m & \rightarrow & \frac{1}{2}\tanh\left(T^{-1}\left(BD-\Theta\right)\right)+\frac{1}{2}\tanh\left(T^{-1}\Theta\right)\label{eq:app_thc}
\end{eqnarray}
because $P_{\mathrm{NL}}\left(m\right)\approx1$ and $C_{\mathrm{NL}}\left(m\right)\approx P_{\mathrm{NL}}\left(-m\right)\approx C_{\mathrm{NL}}\left(-m\right)\approx0$
for $\theta=0$ and $m\rightarrow0$ for $T\rightarrow\infty$. Since
$BD>\Theta$ (App.~\ref{sec:conv}) we have $m>0$ for $T>0$ and
no phase transition occurs (Fig.~\ref{fig:app_thc}, dotted red).

\section{Memory capacity of a stochastic Hopfield network with nonlinear dendritic
branches in the zero temperature limit\label{sec:app_limb}}

We now compute the overlap $m$ in the zero temperature limit $T=0$.
Starting point are the Eqs.~\eqref{eq:app_Fm}, \eqref{eq:app_Fc},
\eqref{eq:app_Fr}, and\eqref{eq:app_Fs}. For $\beta=T^{-1}\rightarrow\infty$,
we may simplify
\begin{eqnarray}
\lim_{\beta\rightarrow\infty}\tanh\left[\pm\beta\left(\bar{F}\left(\pm m\pm\sqrt{\alpha r}z\right)-\Theta\right)\right] & = & \mathrm{sign}\left[\pm\bar{F}\left(\pm m\pm\sqrt{\alpha r}z\right)\mp\Theta\right]\nonumber \\
 & = & \mathrm{sign}\left[m+\sqrt{\alpha r}z\mp\vartheta\right],
\end{eqnarray}
by definition of the effective threshold $\vartheta$ (see Eq.~\eqref{eq:uc}),
$\bar{F}\left(\vartheta\right)=\Theta.$ Using the dominated convergence
theorem, we may compute (Eq.~\eqref{eq:app_Fm})

\begin{eqnarray}
\lim_{\beta\rightarrow\infty}m & = & \frac{1}{2}\int_{-\infty}^{\infty}\frac{dz}{\sqrt{2\pi}}e^{-\frac{1}{2}z^{2}}\mathrm{sign}\left[m+\sqrt{\alpha r}z-\vartheta\right]\nonumber \\
 &  & +\frac{1}{2}\int_{-\infty}^{\infty}\frac{dz}{\sqrt{2\pi}}e^{-\frac{1}{2}z^{2}}\mathrm{sign}\left[m+\sqrt{\alpha r}z+\vartheta\right]\nonumber \\
 & = & \frac{1}{2}\mbox{erf}\left(\frac{m-\vartheta}{\sqrt{2\alpha r}}\right)+\frac{1}{2}\mbox{erf}\left(\frac{m+\vartheta}{\sqrt{2\alpha r}}\right)\label{eq:app_Hm}
\end{eqnarray}
and (Eq.~\eqref{eq:app_Fs})
\begin{eqnarray}
\lim_{\beta\rightarrow\infty}s & = & \frac{1}{2}\int_{-\infty}^{\infty}\frac{dz}{\sqrt{2\pi}}e^{-\frac{1}{2}z^{2}}\mathrm{sign}^{2}\left[m+\sqrt{\alpha r}z-\vartheta\right]\nonumber \\
 &  & +\frac{1}{2}\int_{-\infty}^{\infty}\frac{dz}{\sqrt{2\pi}}e^{-\frac{1}{2}z^{2}}\mathrm{sign}^{2}\left[m+\sqrt{\alpha r}z+\vartheta\right]\nonumber \\
 & = & 1.\label{eq:app_Hs}
\end{eqnarray}
To derive $r$, we further compute $\beta C$ (Eq.~\eqref{eq:app_Fc}).
We use that $\frac{\beta}{2}\left(1-\tanh^{2}\left[\beta x\right]\right)$
approaches the Dirac delta function $\delta\left(x\right)$ for $\beta\rightarrow\infty$,
\begin{eqnarray}
\lim_{\beta\rightarrow\infty}\beta C & = & \int_{-\infty}^{\infty}\frac{dz}{\sqrt{2\pi}}e^{-\frac{1}{2}z^{2}}\delta\left[\bar{F}\left(m+\sqrt{\alpha r}z\right)-\Theta\right]\bar{F}^{'}\left(m+\sqrt{\alpha r}z\right)\nonumber \\
 &  & +\int_{-\infty}^{\infty}\frac{dz}{\sqrt{2\pi}}e^{-\frac{1}{2}z^{2}}\delta\left[-\bar{F}\left(-m-\sqrt{\alpha r}z\right)+\Theta\right]\bar{F}^{'}\left(-m-\sqrt{\alpha r}z\right)\nonumber \\
 & = & \frac{1}{\left|\bar{F}^{'}\left(\vartheta\right)\right|\sqrt{\alpha r}}\int_{-\infty}^{\infty}\frac{dz}{\sqrt{2\pi}}e^{-\frac{1}{2}z^{2}}\delta\left[z+\frac{m-\vartheta}{\sqrt{\alpha r}}\right]\bar{F}^{'}\left(m+\sqrt{\alpha r}z\right)\nonumber \\
 &  & +\frac{1}{\left|\bar{F}^{'}\left(\vartheta\right)\right|\sqrt{\alpha r}}\int_{-\infty}^{\infty}\frac{dz}{\sqrt{2\pi}}e^{-\frac{1}{2}z^{2}}\delta\left[z+\frac{m+\vartheta}{\sqrt{\alpha r}}\right]\bar{F}^{'}\left(-m-\sqrt{\alpha r}z\right)\nonumber \\
 & = & \sqrt{\frac{1}{2\pi\alpha r}}\left[\exp\left(-\frac{\left(m-\vartheta\right)^{2}}{2\alpha r}\right)+\exp\left(-\frac{\left(m+\vartheta\right)^{2}}{2\alpha r}\right)\right].\label{eq:app_Hc}
\end{eqnarray}
In the second line we used $\delta\left(f\left(x\right)\right)=\left|f^{'}\left(x_{0}\right)\right|^{-1}\delta\left(x-x_{0}\right)$
with the (single) root $x_{0}$, $f\left(x_{0}\right)=0$, and again
$\bar{F}\left(\vartheta\right)=\Theta$. There and in the third line
we used that $\bar{F}$ is monotonically increasing, i.e.~$\bar{F}^{'}\geq0$.
From Eqs.~\eqref{eq:app_Fr} and\eqref{eq:app_Hc} we find
\begin{eqnarray}
\sqrt{r} & = & 1+\sqrt{\frac{1}{2\pi\alpha}}\exp\left(-\frac{\left(m-\vartheta\right)^{2}}{2\alpha r}\right)+\sqrt{\frac{1}{2\pi\alpha}}\exp\left(-\frac{\left(m+\vartheta\right)^{2}}{2\alpha r}\right).\label{eq:app_Hr}
\end{eqnarray}
Eqs.~\eqref{eq:app_Hm} and\eqref{eq:app_Hr} provide coupled, implicit
equations for the order parameters $m$ and $r$ which can be solved
numerically. Solutions for varying effective thresholds $\vartheta$
are shown in Fig.~\ref{fig:mag}.

\section{Phase diagrams of associative memory networks of neurons with additive
and non-additive dendritic processing\label{sec:Comparison-of-phase}}

For a further comparison of the memory performance of networks with
arborized neurons with the classical Hopfield model, we compute the
quality of retrieval $m$ in the $\alpha$-$T$-plane.

\begin{figure}[!t]
\noindent \begin{centering}
\includegraphics[width=0.48\textwidth]{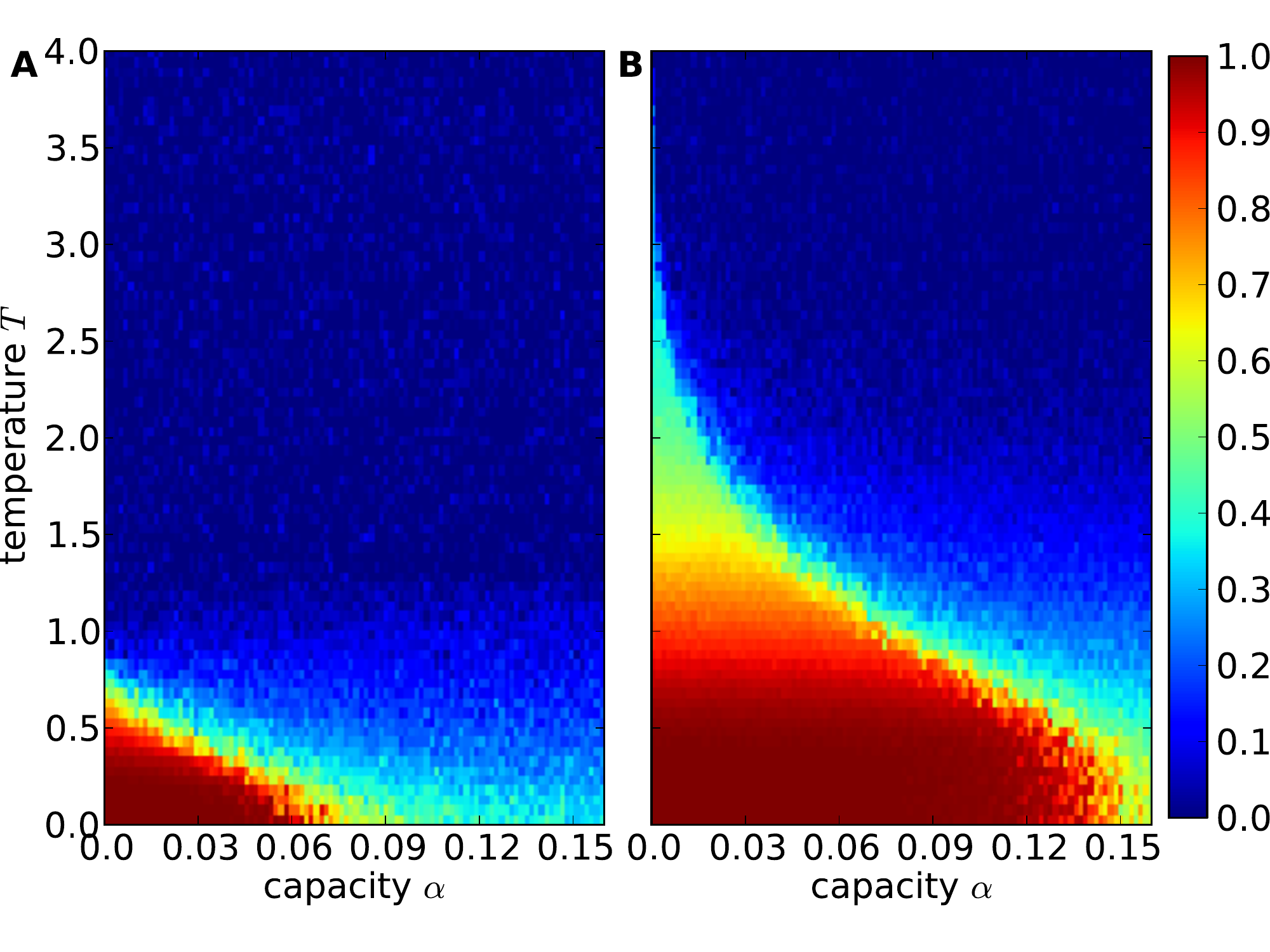}
\par\end{centering}

\protect\caption{\label{fig:phasediags}\textbf{Quality of retrieval in dependence
of capacity $\alpha$ and temperature $T$.} Numerical results for
the overlap $m$, averaged over $20$ realizations. Parameters are
the same as in Fig.~\ref{fig:mag} with $D=0.8$. The network model
with non-additive dendrites displays a larger region of stable memory
retrieval \textbf{(B)} than the original Hopfield network \textbf{(A)}.}
\end{figure}

We found both analytically and numerically that in the limits $\alpha\approx0$
and $T=0$, the critical temperature and capacity of the network with
non-additive dendrites are higher than for the linear model, cf.~Fig.~\ref{fig:mag}.
Complementing these findings, numerical simulations show that the
$\alpha$-$T$-region of successful memory retrieval is larger for
non-additive dendrites (Fig.~\ref{fig:phasediags}B) than for linear
dendrites (Fig.~\ref{fig:phasediags}A).
\end{document}